\documentclass[preprint, showpacs,preprintnumbers,amsmath,amssymb]{revtex4}
\usepackage[dvips]{graphicx}
\usepackage{graphicx}
\usepackage{pstcol}
\usepackage{amsfonts}
\usepackage{bm}
\usepackage{amsmath}
\usepackage{amssymb}
\usepackage{color}
\usepackage[all]{xy}

\def\be{\begin{equation}}
\def\ee{\end{equation}}
\def\bea{\begin{eqnarray}}
\def\eea{\end{eqnarray}}

\begin{document}

\title{Geometrothermodynamic description  of magnetic materials }

\author{Hernando Quevedo}
\email{quevedo@nucleares.unam.mx}
\affiliation{Instituto de Ciencias Nucleares, Universidad Nacional Aut\'onoma de M\'exico, Mexico}
\affiliation{Dipartimento di Fisica and Icra, Universit\`a di Roma “La Sapienza”, Roma, Italy}
\affiliation{Al-Farabi Kazakh National University, Almaty, Kazakhstan}

\author{Mar\'ia N. Quevedo}
\email{maria.quevedo@unimilitar.edu.co} 
\affiliation{Departamento
de Matem\'aticas, Facultad de Ciencias B\'asicas, Universidad
Militar Nueva Granada, Cra 11 No. 101-80, Bogot\'a D.C., Colombia}

\author{Alberto S\'anchez}
\email{asanchez@ciidet.edu.mx} 
\affiliation{Departamento de
Posgrado, CIIDET, {\it AP752}, Quer\'etaro, QRO 76000, MEXICO}

\date{\today}

\begin{abstract}

We perform a statistical and geometrothermodynamic analysis of three different models of magnetic materials, namely,  the translational free model, the spin model, and the mean-field model. First, we derive the fundamental equation for each model, which is then used as input to compute the metrics of the corresponding equilibrium spaces. Analyzing the corresponding geometrothermodynamic curvatures, we conclude that they can be used to describe thermodynamic interaction, stability conditions, and the phase transition structure of the modeled substances. 
In all the cases, we reproduce their well-known behavior close to the Curie temperature. Moreover, in the case of the model with spin, we found a curvature singularity, which corresponds to a novel transition, where a particular response function diverges, indicating the presence of a second-order phase transition, according to Ehrenfest classification. 


\end{abstract}


\maketitle

\section{Introduction}
\label{sec:int}

Any material can be magnetized when placed under the
influence of a magnetic field $\vec{H}$. This is possible because
the molecules and atoms that form the material possess their own 
magnetic properties \cite{Zemansky}. In addition, there are cases
in which, even without the presence of a magnetic field $H$, the
material is magnetized (ferromagnetism); 
this is the case of common magnets. 
In general, ferromagnetism can also be considered as a phase of matter \cite{Hashimoto},
as are the liquid, solid, or gaseous phases of a material.
Ferromagnetism is also temperature-dependent; this means that
there is a critical temperature, called the Curie temperature,
such that above it, the material is paramagnetic, and below it, the
material is ferromagnetic \cite{Berti,Barman}.

From a physical point of view, the phenomenon of magnetism can be described as follows \cite{greiner}. 
All matter consists fundamentally of atoms, and all atoms have electrons whose motion produces electric currents confined in each atom. In turn, the currents produce a magnetic field with a magnetic dipole moment 
 $\vec{\mu}$. If the material is
subject to an external magnetic field $\vec{H}$, then the dipoles
try to align in the direction of the external field. The potential energy
of each dipole is $-\vec{\mu}\cdot \vec{H}$, and all magnetic
moments of the atoms add up to the total magnetic moment (magnetization vector) $\vec{M}$  
\cite{Barman,Mohn}. At a given temperature, the statistical
motion of the dipoles acts against the alignment and, therefore,  from a certain temperature, all the dipoles are statistically
distributed, and the magnetic moments cancel each other so that
the total moment $\vec{M}$ vanishes.

On the other hand, the Ising model \cite{Ising} is one of
the models more frequently used to study the ferromagnetism of
materials. It has also been
applied to many fields of knowledge like biology and neuroscience
\cite{Hopfield,Amit}, economics \cite{Sornette} and sociology
\cite{Kohring,Cajueiro}.  Unfortunately, 
solutions of this model in physics can only be studied in one or two
dimensions \cite{Turban,Messager,Unnar,Kedkanok,Vecsei}, and these
solutions are usually very complicated.
 However, it is possible to obtain approximate solutions in
any number of dimensions using mean-field theory, which is based upon the assumption that the system's thermal fluctuations are small
and can, therefore, be neglected to a certain extent. In this way,
we can study a system of particles that only interact with an effective
mean field, which captures the average behavior of the particles
around it \cite{Dalton}. In the mean-field approximation, the Hamiltonian
becomes simple, which allows us to study the behavior of
complicated many-body systems that cannot be solved exactly. 
Therefore, it is very important to investigate the physical and mathematical structure of magnetic models. In this work, we propose to investigate magnetic models from the point of view of differential geometry, as formulated within the framework of geometrothermodynamics.

Furthermore, the thermodynamics of magnetic systems has been the subject of
many studies, in particular, because there is a great variety of situations where
this subject can be applied \cite{castellanos, Barrett, Berez, Apol,
Moore}. In this connection, the fundamental equation and the
thermodynamic potentials play an important role because, if one of
them is known, all the thermodynamic properties of the system can
be obtained \cite{callen}. The fundamental equation and the
thermodynamic potentials can be obtained in two different ways.
The first corresponds to the case where all state equations of the
thermodynamic systems are known. The second way is statistical mechanics, using the partition function. In the
present paper, we will work with the second possibility in order
to get the corresponding thermodynamic potentials necessary to
study the thermodynamic and the geometrothermodynamic background of magnetic systems.

On the other hand, the important relationship between differential geometry and physics is well known, particularly in thermodynamics. Since the pioneer works by  Gibbs and Charatheodory
 \cite{Gibbs,Charatheodory}, a variety of studies in this direction have been performed. We highlight the contributions by Rao
\cite{Rao} in statistical physics, information theory, and
thermodynamics \cite{Amari}, the works by Weinhold, Ruppeiner, and Mrugala
\cite{Mrugala1,Mrugala2,Weinhold,Ruppeiner1,Ruppeiner2}, who set the fundamentals of thermodynamic geometry by postulating and investigating the Hessians of 
the internal energy and entropy as Riemannian metrics for the space of equilibrium states, which is an $n$-dimensional manifold with points representing the equilibrium states of the system, and $n$ is the number of its thermodynamic degrees of freedom. The integer $n$ corresponds also to the number of independent variables that need
to be specified to reproduce the experimental realization of the
system. The independent variables are the coordinates of the
manifold, and the curvature of the equilibrium space is related to
 properties of the system such as thermodynamic interactions and phase transitions. 
The most recent formalism proposed to
develop  a geometric representation of  thermodynamic properties
is called geometrothermodynamic (GTD) \cite{quevedo}. 
GTD utilizes concepts of contact geometry and differential geometry in two
Riemannian manifolds; one of them is the standard equilibrium space and the second one is the thermodynamic phase space, having as its principal ingredient Legendre
invariance, thereby ensuring that the properties of a system are
invariant with respect to changes of the thermodynamic
potential, i.e., invariant with respect to Legendre
transformations. GTD has been applied to many thermodynamic
systems, leading to consistent results
\cite{quevedo1,quevedo3,quevedo4,quevedo5,QSI,QSII}.

In this work,  we investigate the 
thermodynamics of magnetic materials using GTD. We will see that the thermodynamic properties of magnetic materials can be described by the geometric properties of the equilibrium space, in particular,  curvature represents thermodynamic interaction and curvature singularities correspond to phase transitions. 
We will study three different models, namely, 
the non-translational model, the magnetic dipole moment with spin, 
 and the mean-field model. We use the formalism of statistical physics to derive the fundamental equation of each model and to compute the GTD metrics for the corresponding equilibrium spaces.  We will show that the translational free model is characterized by the presence of thermodynamic interaction that does not lead to phase transitions. In the case of the model of dipole moments with spin, we will see that there is a curvature singularity representing a novel phase transition, which we interpret as due to a divergence in a particular response function.
The mean-field model exhibits curvature singularities that we interpret as related to the violation of the equilibrium condition and the presence of phase transitions. We thus reproduce, at the level of the geometrothermodynamic curvature, the behavior of magnetic materials with respect to the Curie temperature and find a novel phase transition in materials composed of magnetic dipoles with spin. 

The paper is organized as follows: In Sec. 
\ref{sec:models}, we use the standard formalism of statistical physics to derive the thermodynamic potentials of the three models.
 In Sec. \ref{sec:gtd}, we introduce the main
concepts of GTD. In Sec. \ref{sec:gtdmag}, we apply the formalism of GTD to each model and compute the main geometric properties of the corresponding equilibrium spaces. 
Finally, Sec. \ref{sec:con}
is devoted to
conclusions.

\section{Thermodynamics of magnetic materials}
\label{sec:models}

This section is dedicated to the study of three models for magnetic materials from the point of view of statistical physics. The aim is to derive the fundamental thermodynamic equation for each model, from which we will extract the complete thermodynamic and geometrothermodynamic information. 

\subsection{Magnetic model without translation}

First of all, let us consider a system of $N$ magnetic dipoles that
freely revolve and whose translational motions are neglected. We
consider classical dipoles, which can assume all possible
orientations. In this model, the energy of such a material is assumed to be determined by 
\bea \label{energy1} E=-\sum_{i=1}^{N}\vec{\mu}_i \cdot
\vec{H}\,,\eea 
where $\vec{\mu}_i$ represents the $i-$th magnetic
dipole moment. Assuming that the homogeneous field $\vec{H}$
points in the $z$--direction, the partition function $Z$ associated
with this model has the form \cite{greiner}
\bea \label{partition1} Z(T,H)=\frac{4\pi k_{{}_B} T}{\mu H}
\sinh{\left(\frac{\mu H}{k_{{}_B} T} \right)}\,,\eea where $k_{{}_B}$, $T$
and $H$ represent the Boltzmann constant, the temperature, and the
magnetic field, respectively. We also use the notation 
$\vec{\mu}_i \cdot \vec{H}=\mu_{zi} H_z=\mu H\cos\theta_i$. Then,
according to statistical mechanics, the Helmholtz free energy $F$,
the internal energy $U$, the entropy $S$, and the magnetization $M$
are given by the expressions \cite{greiner}
\bea \label{HFE} F&=&-Nk_{{}_B}T\ln{Z}=-Nk_{{}_B}T\ln{\Bigg[4\pi
\frac{\sinh{(\beta \mu H)}}{\beta \mu H}\Bigg]}\,,\\
\label{Ienergy} U&=&-\frac{\partial [\ln Z]}{\partial
\beta}=NH\mu\Bigg[\frac{1}{\beta \mu H}-\coth{(\beta \mu H)}
\Bigg]\,,\\ \label{Entropy} S&=&k_{{}_B} \beta \frac{\partial
F}{\partial \beta}=Nk_{{}_B}\mu H\Bigg[\coth{(\beta \mu
H)}-\frac{1}{{\beta \mu H}} \Bigg]\,,\\ \label{Magnetization}
M&=&-\frac{\partial F}{\partial H}=N \mu \left[\coth{(\beta \mu
H)}-\frac{1}{\beta \mu H} \right]
\label{mag1}
\,,\eea 
with
$\beta=\frac{1}{k_{{}_B} T}$. 

\begin{figure}[h] 
\includegraphics[scale=0.33]{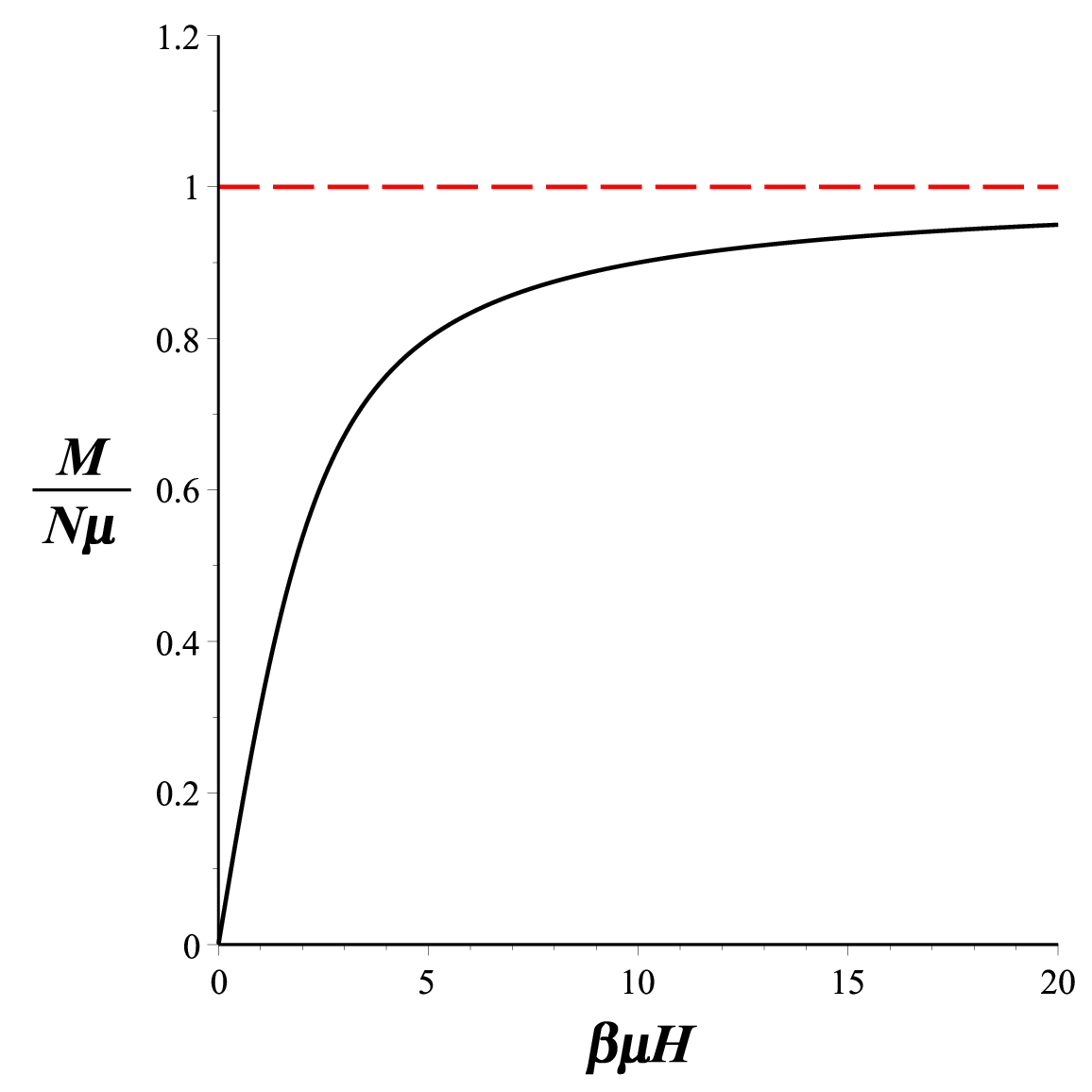}
\caption{Magnetization as a function of the temperature $T$
$(\beta \mu H= \frac{\mu H}{k_{{}_B} T} )$.}
\label{fig1}
\end{figure}

The thermodynamic properties of this model can be derived from any of the above potentials. Let us consider, in particular, the behavior of the
magnetization density $\frac{M}{N\mu}$ in terms of the temperature. This  is
shown in Fig. \ref{fig1}, where 
we can see that for finite values ($\beta \mu
H<<1$) the magnetization behaves linearly with the magnetic field,
and the magnetic dipole moments are of the order of the Bohr magneton
$\mu_B=\frac{e\hbar}{2m_e}$.

Using a Taylor decomposition, the magnetization  (\ref{Magnetization}) in the limit $\beta \mu H << 1$ can be written as
\bea \label{MaglowT} \frac{M}{N\mu}\approx
 \frac{\beta \mu H}{3}= \frac{\mu}{3k_{{}_B}T}H\,, 
 \eea 
 where the
 proportionality constant $\frac{\mu}{3k_{{}_B}T}$ is called magnetic susceptibility and denoted by 
 \bea \label{suseptibilidad} \chi=\frac{C}{T}\,,\eea 
 where
$C=\frac{\mu}{3k_{{}_B}}$. This is the Curie law, and $C$
represents Curie's constant.
In the case $\beta \mu H>>1$, i.e., for 
 $T\longrightarrow 0$, the saturation region is
reached, and nearly all dipoles are aligned in the field direction.

\begin{figure}[h] 
\includegraphics[scale=0.33]{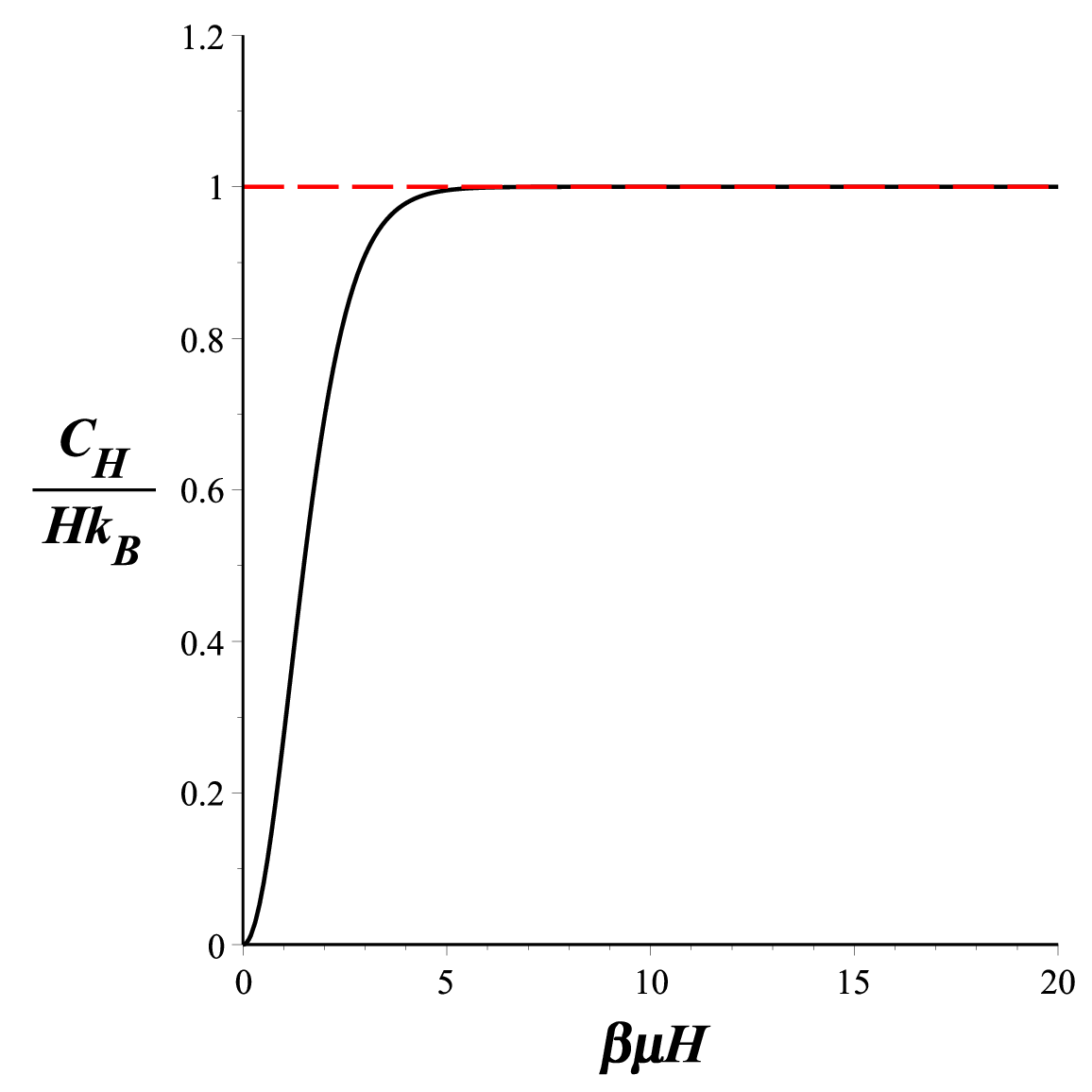}
\caption{Heat capacity at constant $H$.}
\label{fig2}
\end{figure}

The heat capacity at a constant magnetic field  can be computed by
the relationship \cite{greiner}
\bea \label{heatcapacity} C_{{}_H}=-k_{{}_B} \beta^2
\frac{\partial U}{\partial \beta}\,.\eea
Using the relation (\ref{Ienergy}) and Taylor's series for 
$\sinh(\beta \mu H)$, it is possible to write the next expression
for the heat capacity
\bea \label{heatcapacity2} C_H=Nk_{{}_B}\Bigg[1-\Bigg[\frac{\beta
\mu H}{\beta \mu H+\frac{(\beta \mu H)^3}{3!}+\frac{(\beta \mu
H)^5}{5!}+\cdots+\frac{(\beta \mu H)^{2n+1}}{(2n+1)!}}\Bigg]^2 \Bigg]\,,\eea
with $n=0,1,2,\cdots$.
For high temperatures, $\beta \mu H \longrightarrow 0$, Eq.  (\ref{heatcapacity2}) tells us that the heat capacity
goes to zero. This is the case for systems whose energy has an
upper limit, indicating that the magnetic dipoles do
not yield any contribution to the heat capacity; Fig. \ref{fig2} shows
this behavior.

\subsection{Magnetic model with spin}

In this model, the magnetic dipole moment is an operator
$\hat{\mu}$ which is defined as
\bea \label{para14} \hat{\mu}=\Big[ g_l \hat{l}+g_s \hat{s}
\Big]\mu_B\,,\eea 
where $\hat{l}$ and $\hat{s}$ are the angular
momentum and spin operators, respectively. The factor
$\mu_B=\frac{e\hbar}{2m_e c}$ is Bohr's magneton. The parameter
$g$ is called the gyromagnetic factor and takes the following values:
$g_l=1$ y $g_s\approx 2$ for electrons.

In a system where the angular momentum is conserved, as is the
case of atoms, the magnetic dipole moment $\vec{\mu}$ will
presses around  $\vec{j}$, and on average only the projection of
$\vec{\mu}$ onto $\vec{j}$ remains constant. According to quantum
mechanics, this magnetic dipole moment has the form,
\bea \label{para17} \vec{\mu}_p=\left[
\frac{3}{2}+\frac{s(s+1)-l(l+1)}{2j(j+1)}\right]\mu_B\vec{j}=g\mu_B\vec{j}\,,\eea
where $s(s+1)$, $l(l+1)$ and $j(j+1)$ are the eigenvalues of
operators $\hat{s}^2$, $\hat{l}^2$, and $\hat{j}^2$, respectively.
Eq. (\ref{para17}) remains valid also for the total
magnetic moment of all electrons if the total quantities $S$, $L$,
and $J$ are considered. Then, the eigenvalues of the energy of a
dipole in a magnetic field would be:
\bea \label{para18} E=-\vec{\mu}_p\cdot \vec{H}\,.\eea
Using the relationship (\ref{para17}), and considering a magnetic
field $\vec{H}$ in the $z$-direction, Eq.  (\ref{para18})
takes the form
\bea \label{para19} E=-g\mu_B H j_z=-g\mu_B H m\,,\eea 
where
$m=-j,-j+1,...,+j$ is the component of $\vec{j}$ in the magnetic
field direction. For a system of $N$ dipoles, the partition
function has the following form \cite{greiner}
 \bea \label{para20}
Z(T,H,N)=\sum_{m_1, m_2,...,m_N=-j}^{+j}\exp{\Big[\beta g \mu_B
H\sum_{i=1}^N m_i\Big]}\,,\eea 
where, for all $N$ dipoles, the
sum extends over all possible orientations $m_i$. Considering
the case $j=-\frac{1}{2}$ with $g=2$, the dipole only has two
possible energy values:
\bea \label{para23} E=-2\mu_B H m\, ,\quad \quad m=-\frac{1}{2},
+\frac{1}{2}\,.\eea 
Then, the partition function takes the simple form
\bea \label{para25} Z(T,H,N)=\Big[ 2\cosh{(\beta \mu_B
H)}\Big]^N\,.\eea
Furthermore, from the partition function, we compute the
corresponding Helmholtz free energy $F$
\bea \label{para26}
F(T,H,N)=-k_{{}_B}T\ln{Z(T,H,N)}=-Nk_{{}_B}T\ln{\Big[
2\cosh{(\beta \mu_{{}_B} H)}\Big]}\,,\eea 
which corresponds to a special case of the
one-dimensional Ising model \cite{Ising}.

\begin{figure}[h] 
\includegraphics[scale=0.33]{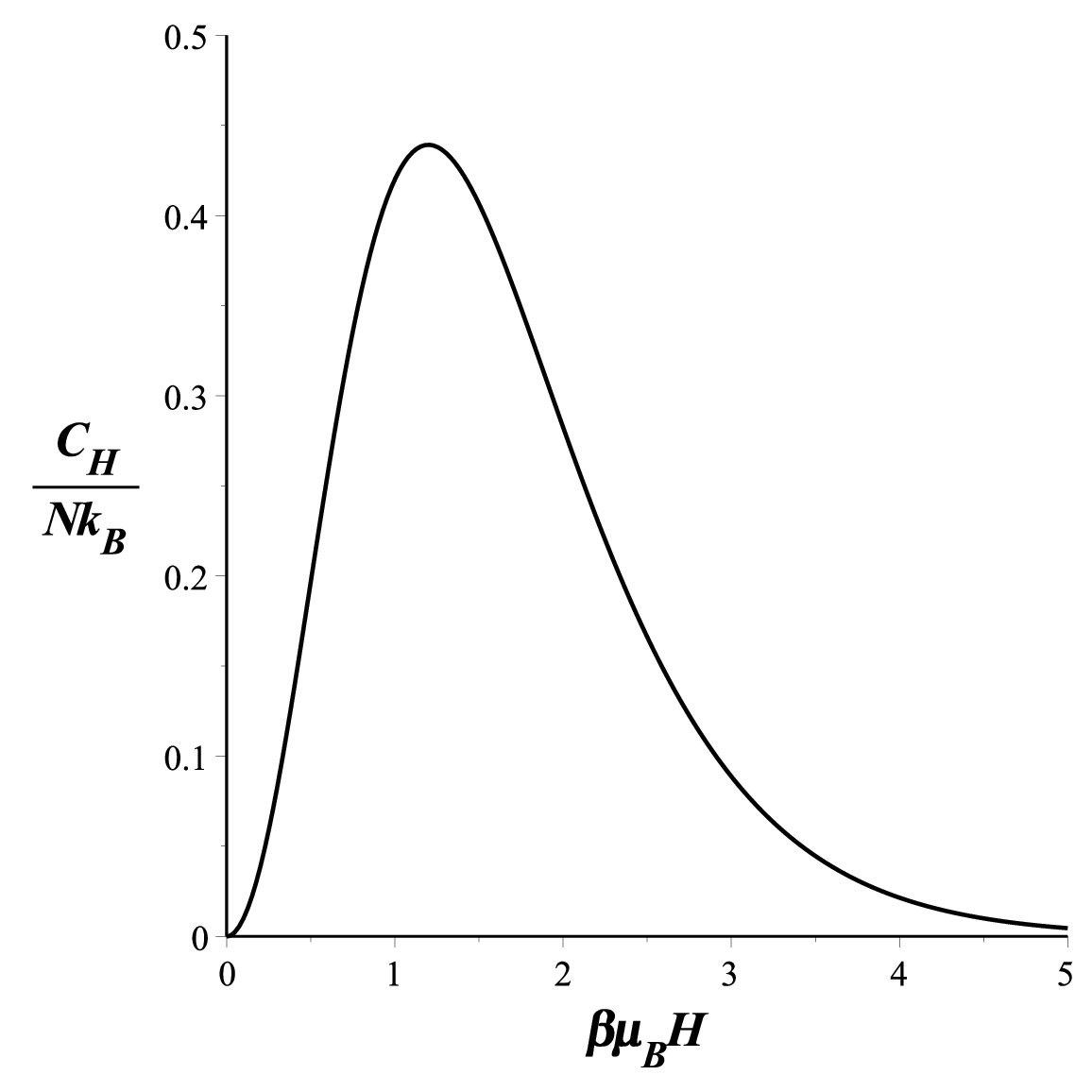}
\caption{Heat capacity at constant $H$.}
\label{fig3}
\end{figure}

Again, using the partition function (\ref{para25}) and the
Helmholtz free energy (\ref{para26}), it is possible to calculate
the corresponding internal energy $U$, the entropy $S$, and the
magnetization $M$. These quantities have the form:
\bea \label{para32}  U&=&-N\mu_{{}_B} H\tanh{(\beta \mu_{{}_B} H)}\,,\\
\label{para31} S&=& Nk\Bigg\{\ln{\Big[2\cosh{(\beta \mu_{{}_B}
H)}\Big]} -\beta \mu_{{}_B} H \tanh{(\beta \mu_{{}_B} H)} \Bigg\}\,,\\
\label{para30} M&=& N\mu_{{}_B}\tanh{(\beta \mu_{{}_B} H)}\,.\eea
Finally, using the defining equation (\ref{heatcapacity}) we get the heat
capacity for  this model as
\bea \label{para33} C_H=Nk_{{}_B}\Big[\beta \mu_{{}_B} H
\Big]^2\cosh^{-2}{(\beta \mu_{{}_B} H)}\,,\eea 
whose behavior is depicted in Fig. \ref{fig3}.
We see that the heat capacity is represented by a monotonic function in terms of the inverse of the temperature. No divergences are present, indicating that no phase transitions exist.

\subsection{Mean-field magnetic model}

In the magnetic model of the last section, the electron's spin was
considered, but the interaction between spins was not taken into
account. The mean-field magnetic model considers that each
electron's spin interacts with its neighbors. Therefore, if
each spin $s_i$ has $q$ neighbors, in the mean-field model, the
intensity of the magnetic field is replaced by the quantity
\bea \label{campototalmedio} H+(N-1)^{-1}qI\sum_{j\neq i}
s_j\,,\eea 
where $I$ is a positive (ferromagnetic) coupling
constant. Now the sum is over the $(N-1)$ sites $j$ different of
$i$. Accordingly, the energy will take the form
\bea \label{energia2} E=-\frac{1}{2}qI\frac{m^2-N}{N-1}-Hm\,,\eea
where $m=\sum_{i=1}^N s_i$. The sum over the spins in the
partition function can be replaced by the sum of the allowed
values of $m$, weighted by the number of spin configurations for
each value. If $r$  spins are oriented downward (their value is
$-1$), then, $N-r$ are upward (their value is $+1$), the 
equation for $m$ can be written as $ m=N-2r$, and there are
$\frac{N!}{r!(N-r)!}$ arrays of spins. Thus,  the partition function
for this model can be written as \cite{baxter} 
\bea \label{particionmedio} Z=\sum_{r=0}^N
{\frac{N!}{r!(N-r)!}}\exp{\Big\{ \frac{\beta q I [
(N-2r)^2-N]}{2(N-1)} +\beta H(N-2r)\Big\}}\,.\eea

According to statistical mechanics, the magnetization will have
the form
\bea \label{magnetizacionnmedio} M=\frac{\sum_{r=0}^N
{(1-\frac{2r}{N})}c_r}{\sum_{r=0}^N c_r}\,\eea where
\bea \label{coeficientes} c_r=\frac{N!}{r!(N-r)!}\exp{\Big \{
\frac{\beta q I [ (N-2r)^2-N]}{2(N-1)} +\beta H(N-2r)\Big\}
}\,.\eea 
In order to analyze the quantities $c_r$, we consider the related
quantity
\bea \label{coeficientes2}
d_r=\frac{c_{r+1}}{c_r}=\frac{N-r}{r+1}\exp{\Big \{ -2\frac{\beta
q I [ N-2r-1]}{(N-1)} -2\beta H\Big\} }\,.\eea
As $r$ increases from $0$ to $N - 1$, the 
right-hand side of
(\ref{coeficientes2}) increases from large values ($\sim$ $N$) to
small values ($\sim$ $N^{-1}$). Provided $\beta q I$ is not too
large, this decrease must be monotonic, and  there must be a
single integer such that
\bea \label{condiciones1} d_r &>&1  \quad\quad for \quad\quad
r=0\cdots , r_0-1\,,\\ \label{condiciones2} d_{r_0}&\leq & 1\,,\\
\label{condiciones3} d_r &<& 1  \quad\quad for \quad\quad
r=r_0+1\cdots , N-1\,.\eea 
Since $c_{r+1}=d_r c_r$, it follows
that $c_r$ increases as $r$ goes from $0$ to $r_0$, decreases as
$r$ goes from $r_0+1$ to $N$, and that $c_{r_0}$ is the largest
 $c_r$.

Considering that $N$ and $r$ are very large, we have that
\bea \label{aprox1} r+1\approx r\,, \quad \quad \frac{
N-2r-1}{(N-1)} \approx 1-\frac{2r}{N}\,,\eea
and Eq. (\ref{coeficientes2}) takes the form
\bea \label{coeficientes3} d_r (x)=\frac{N-r}{r}\exp{\Big \{
-2\beta q I \Big( 1-\frac{2r}{N}\Big) -2\beta H\Big\} }\,,\eea
with $-1<1-\frac{2r}{N}<1$.

In this approximation, it is not difficult to show that the
magnetization becomes
\bea \label{coeficientes13} M=\tanh{[ \beta( q I M + H)\big]
}\,.\eea
Considering that $M=- \frac{\partial F}{\partial H}$ and using the
chain rule, we obtain the Helmholtz free energy for this magnetic
model as
\bea \label{valormediomagnetico11} -\beta F=\frac{1}{2}\ln\left(
\frac{4}{1-M^2}\right) -\frac{1}{2}\beta q I M^2\,.\eea

In order to study the phase transitions of this model, we get the
magnetic field $H$ from the relation (\ref{coeficientes13})
\bea \label{coeficientes14} H=\frac{ {\rm artanh}{(M)}}{\beta} - q I M
\,.\eea

\begin{figure}[h] \centering \includegraphics[angle=360, scale=0.25]{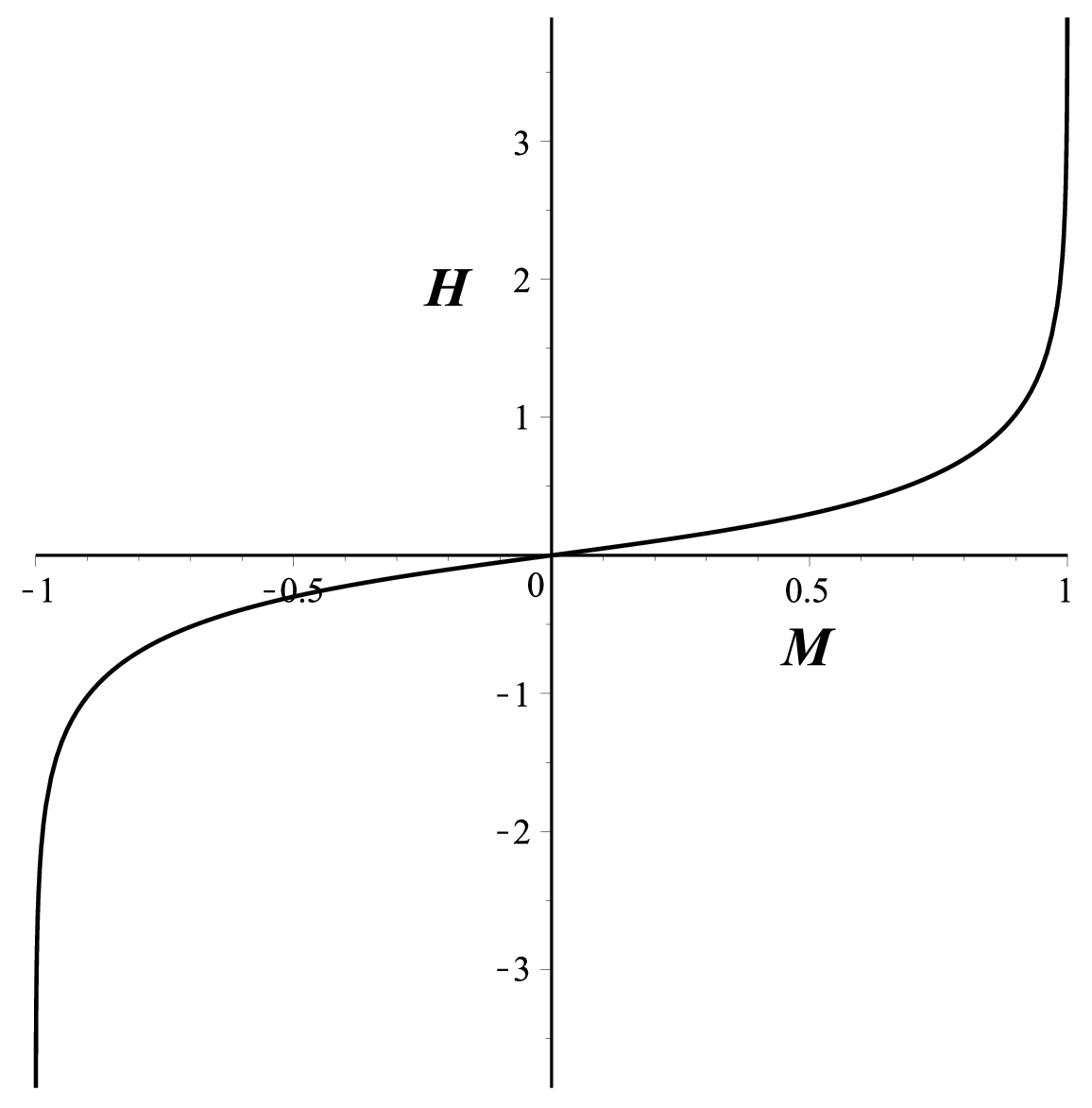}\includegraphics[angle=360, scale=0.25]{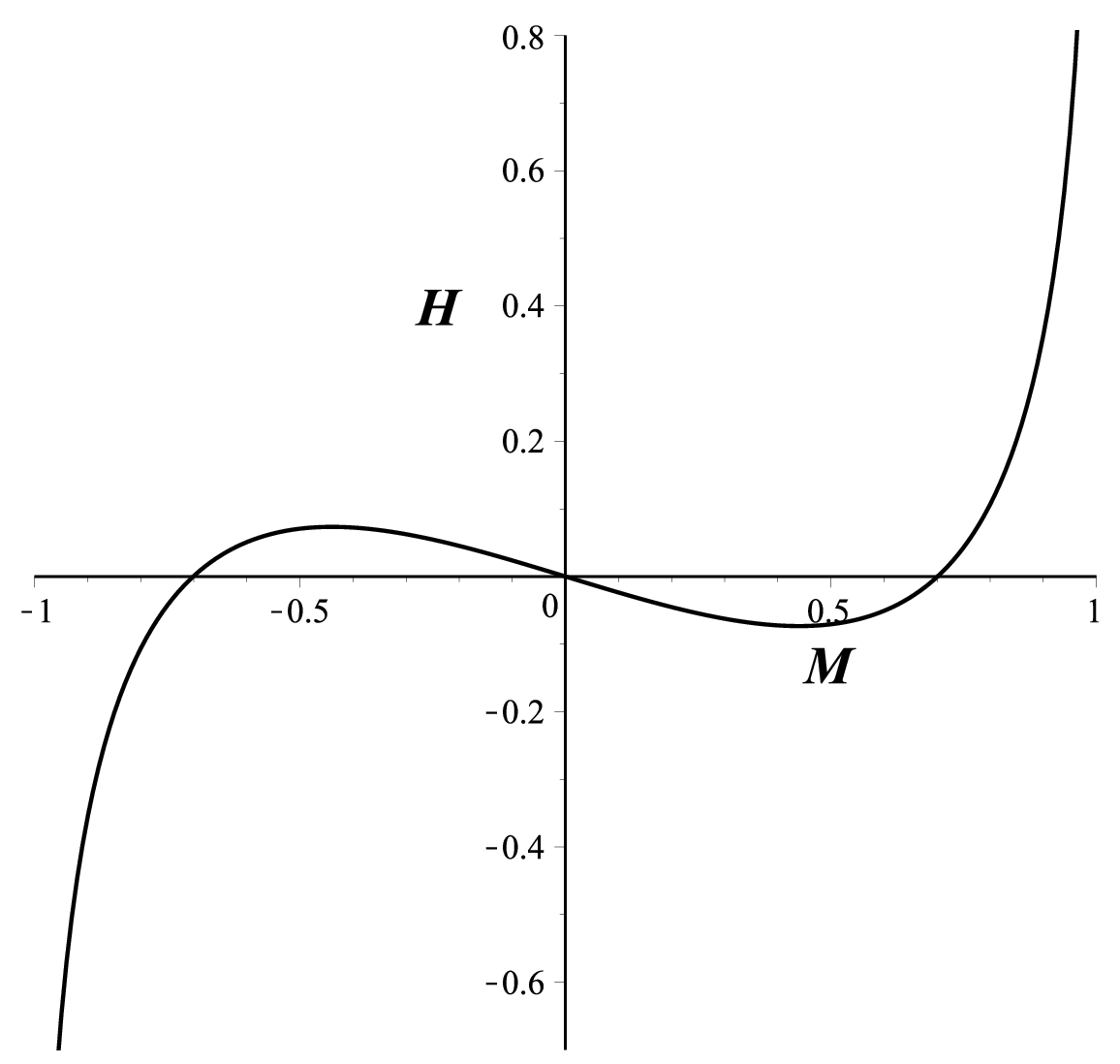}
\caption{Magnetic field $H$ as a function of the magnetization
$M$. Here, $ qI\beta <1$ (left), $qI\beta >1$ (right) }
\label{fig4}
\end{figure}

If $qI\beta <1$, at high temperatures there is no
spontaneous magnetization (remnant). This behavior is shown in the
left graph of Fig.\ref{fig4}.

In the case $qI\beta >1$, the behavior of the magnetic field will
be as shown in the right graph of Fig.\ref{fig4}. However, this graph tells us that for a sufficiently small
magnetic field, there are three possible values of magnetization,
whereas the magnetization $M$, defined by $M=-\frac{\partial
F}{\partial H}$, implies a single-valued function of $H$, which
contradicts the statements preceding Eq.(\ref{condiciones1}).

\begin{figure} [h] \centering \includegraphics[angle=360, scale=0.22]{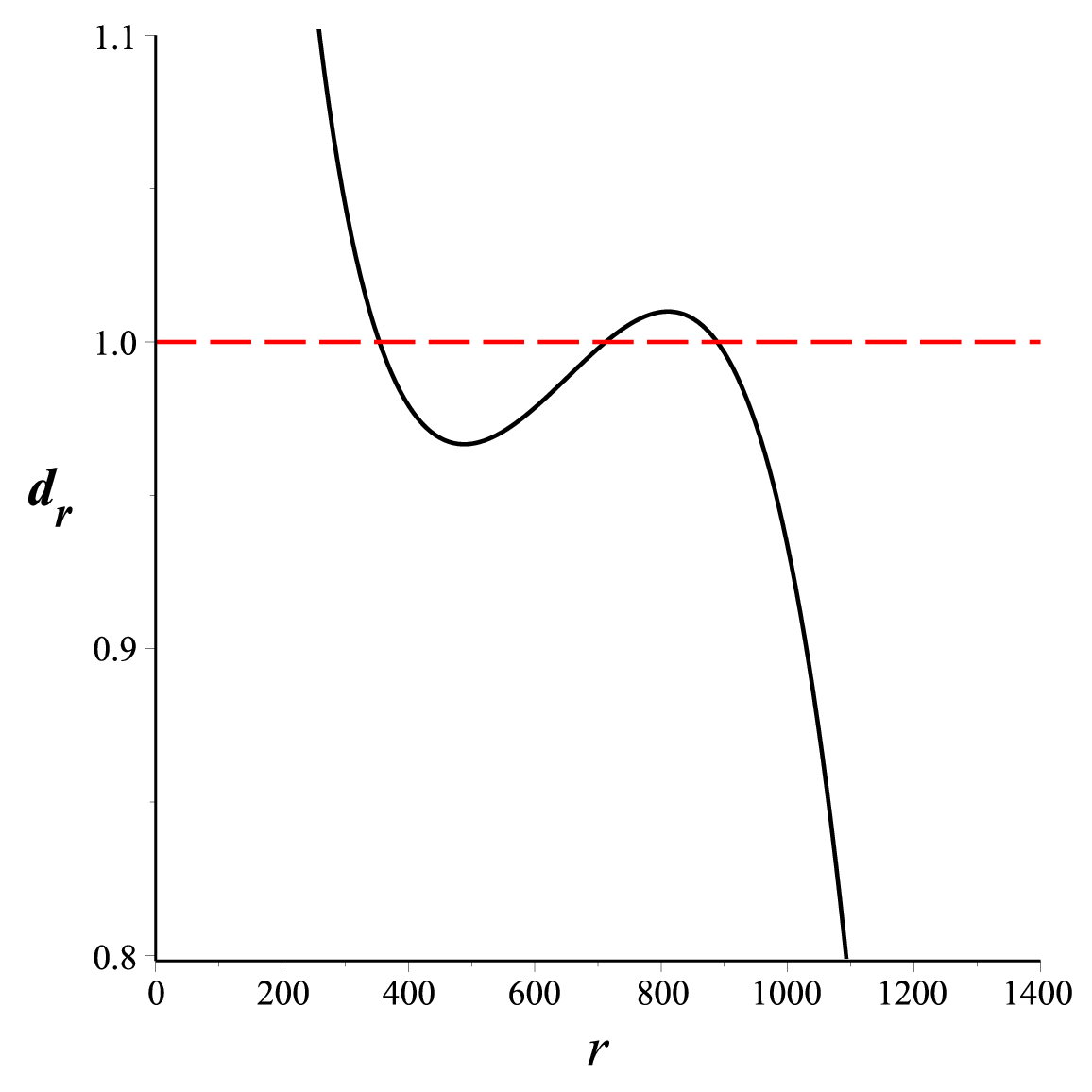}\includegraphics[angle=360, scale=0.22]{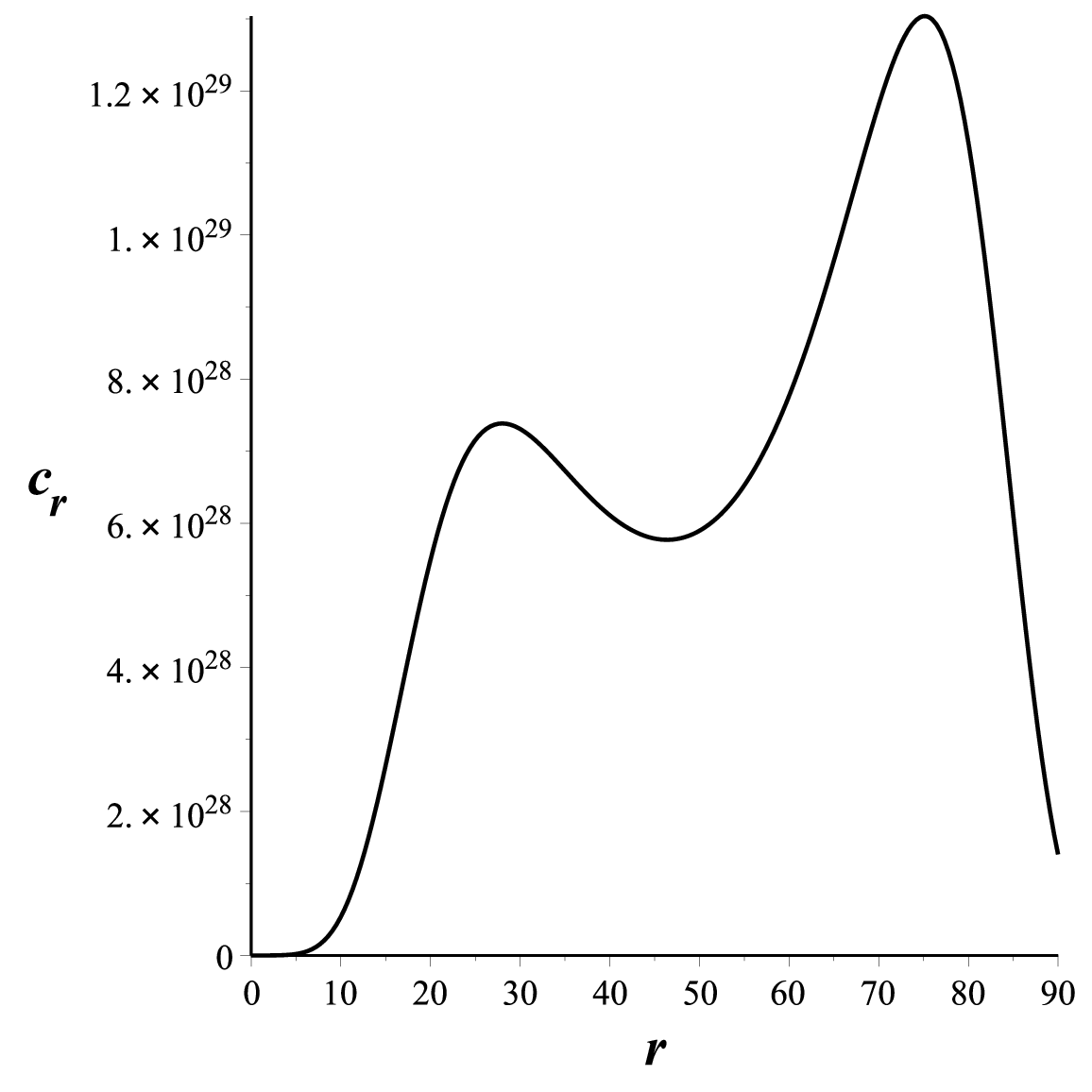}
\caption{  Left graph:  $d_r$ as a function of $r$ for  $\beta H = 0.006$
and $N=1300$. Right graph:
$c_r$ as a function of $r$ with $\beta H = 0.006 $ and $N=100$ .}
\label{fig5}
\end{figure}

In this case, $C_r$  has two maxima and a minimum, as shown in the
graph on the right-hand side of Fig.\ref{fig5}. These correspond to the three
solutions for $M$ of the Eq.(\ref{coeficientes13}). If $H$ is
positive (negative), then the left-hand (right-hand) peak is
greater.

\begin{figure} [ht] \centering \includegraphics[angle=360, scale=0.25]{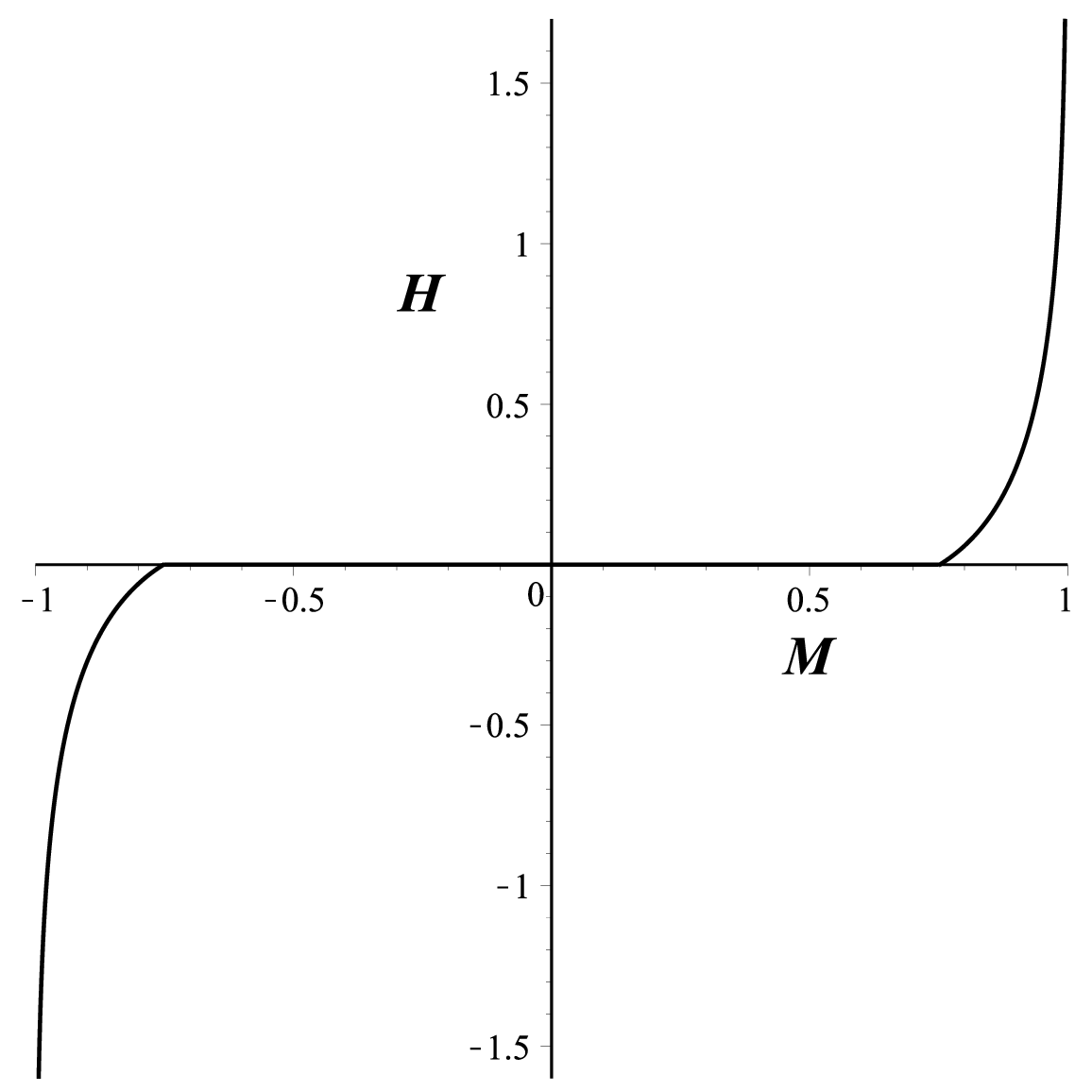}
\caption{ Behavior of the equation (\ref{coeficientes14}) for $
qI\beta <1$ without the non-essential solutions.}
\label{fig6}
\end{figure}

Therefore, when $H$ is positive, we must choose the solution with
the smallest value of $r_0$, i.e., the largest value of $M$.
Conversely, if $H$ is negative, the graph on the
right-hand side of Fig. \ref{fig4} becomes the graph of Fig. \ref{fig6}. There is a
spontaneous magnetization (remnant magnetization) $M_0$, which
corresponds to $H=0$, given by Eq.(\ref{coeficientes13})
as
\bea \label{remanente} M_0=\tanh{ (\beta q I M_0)
}\,,\eea  
as far as $qI\beta >1$. Therefore, the mean-field
model has a ferromagnetic phase transition at temperatures below
the Curie temperature
\bea \label{temperaturacurie2} T_c=\frac{qI}{k_{{}_B}}\,.\eea

\section{Review of geometrothermodynamics}
\label{sec:gtd}

Geometrothermodynamics \cite{quevedo} is a formalism
that accurately describes the thermodynamic properties of
physical systems in terms of concepts of differential geometry, taking into account the symmetry properties of classical thermodynamics. Indeed, the formulation of GTD uses the
Legendre invariance of classical thermodynamics as a fundamental ingredient, which allows us to obtain consistent results and to describe in an invariant geometric way  phase transitions and thermodynamic
interactions \cite{QSI,QSII,quevedo4}.

In order to formulate Legendre invariance, the starting point of
GTD is a $(2n + 1)$-dimensional  space $\mathcal{T}$ with
coordinates $Z^A=\{ \Phi, E^a, I_a \}$, where $A = 0, 1, . . . ,
2n $, which allows us to represent Legendre transformations as coordinate transformations in $\mathcal{T}$ as follows \cite{Arnold}:
$  \{ Z^A\} \rightarrow  \{ \tilde{Z}^A\}=\{ \tilde{\Phi},
\tilde{E}^a, \tilde{I}_a \}$ with 
$\Phi=\tilde{\Phi}-\tilde{E}^k \tilde{I}_k, 
\  E^i=-\tilde{I}^i, \  E^j=\tilde{E}^j,$ and $
\ I^i=\tilde{E}^i, \ I^j=\tilde{I}^j$, 
where $i \cup j$ is any disjoint
decomposition of the set of indices $\{1, . . . , n\}, k, l =1,
\dots, i$. For $i = \emptyset$, we obtain the identity
transformation and for $i = \{1, . . . , n\}$ a total Legendre
transformation.

The main characteristic of the  space $\mathcal{T}$ is that,
according to the Darboux theorem, there exists a
canonical 1-form $\Theta = d\Phi - I_a dE^a$ , which satisfies the
condition $ \Theta \wedge (d\Theta)^n\neq 0$, where  $\wedge$ denotes the exterior product, and $d$
the exterior derivative. $\Theta$ is called contact 1-form and 
is Legendre invariant in the sense that under a Legendre
transformation  it behaves as
$\Theta  \rightarrow \tilde{\Theta} = d\tilde{\Phi} -
\tilde{I}_a d\tilde{E}^a$, implying that the functional
dependence of $\Theta$ remains unchanged. Moreover, the phase space is also equipped with a Riemannian metric $G_{AB}$, which is demanded to be Legendre invariant. This condition is satisfied by three different classes of metrics whose line elements can be written as \cite{QuevedoMN5}
\be
G^{^{I}}=  (d\Phi - I_a d E^a)^2 + (\xi_{ab} E^a I^b) (\delta_{cd} dE^c dI^d) \ ,
\label{GI}
\ee
\be 
G^{^{II}}= (d\Phi - I_a d E^a)^2 + (\xi_{ab} E^a I^b) (\eta_{cd} dE^c dI^d) \ ,
\label{GII}
\ee
\be	
\label{GIII}
G^{{III}}  =(d\Phi - I_a d E^a)^2  + \sum_{a=1}^n \xi_a (E^a I^a)^{2k+1}   d E^a   d I^a \ ,
\ee
where $\eta_{ab}= {\rm diag}(-1,1,\cdots,1)$, $\xi_a$ are real constants, $\xi_{ab}$ is a diagonal $n\times n$ real matrix, and $k$ is an integer. 
 The set $(\mathcal{T}\,,\Theta\,,G)$ defines a
Riemannian contact manifold and is called the thermodynamic phase
space of GTD. 

In GTD, the equilibrium space $\mathcal{E}$ is defined as a subspace of $\mathcal{T}$ by means of a smooth embedding map $\varphi: \mathcal{E}\to \mathcal{T}$ such that $\varphi^*(\Theta)=0$, where $\varphi^*$ is the corresponding pullback. The line element $G= G_{AB} dZ^ A d Z^B$ on $\mathcal{T}$ induces a line element $g=g_{ab}dE^a dE^b$ on $\mathcal{E}$ by means of the pullback, i.e., $\varphi^*(G)=g$, where for concreteness we choose $E^a $ as the coordinates of $\mathcal{E}$.
Then, from Eqs.(\ref{GI}), (\ref{GII}), and (\ref{GIII}), we obtain
\be
g^{{I}} =  \sum_{a,b,c=1}^n \left( \beta_c E^c \frac{\partial\Phi}{\partial E^c} \right)  
\frac{\partial^2\Phi}{\partial E^a \partial E^b}  dE^ a d E^ b ,
\label{gdownI}
\ee
\be
g^{II} =   
 \sum_{a,b,c,d=1}^n \left( \beta_c E^c \frac{\partial\Phi}{\partial E^c} \right) 
 \eta_a^{\ d}
\frac{\partial^2\Phi}{\partial E^b \partial E^d} dE^ a d E^ b   ,
\label{gdownII}
\ee
\be
g^{{III}} = \sum_{a,b=1} ^n \left( \beta_a  E^a \frac{\partial\Phi}{\partial E^a}\right)
 \frac{\partial ^2 \Phi}{\partial E^a \partial E^b}
dE^a dE^b \ ,
\label{gdownIII}
\ee
respectively, 
where $\eta_a^{\ c}={\rm diag}(-1,1,\cdots,1)$. To obtain the above results, we have 
chosen the free constants as $\xi_a=\beta_a$ and $\xi_{ab} = {\rm diag}(\beta_1,\cdots,\beta_n)$, where the parameters $\beta_a$ are defined from the property that the fundamental equation $\Phi=\Phi(E^a)$ is a 
quasi-homogeneous function, i.e., $\Phi(\lambda^{\beta_a}E^a)=\beta_\Phi \Phi(E^a)$ for real constants $\lambda$, $\beta_a$, and $\beta_\Phi$. Moreover, the constant $k$ has been chosen as $k=0$, which follows from the condition that all three metrics can be applied to the same thermodynamic system simultaneously and lead to compatible results.  Furthermore, if the Euler relation, $\sum_a \beta_a E^ a\frac{\partial \Phi}{\partial E^ a} = \beta_\Phi \Phi$, is satisfied, the conformal factor in front of the components $g^I_{ab} $ and $g^{II}_{ab}$ can be replaced by $\beta_\Phi \Phi$, which simplifies calculations.

 In particular, for the case $n=2$ the resulting line elements can be written as
\bea
\label{gI2D} 
g^{I}& = & \Sigma \left[\Phi_{,11} (d E^1)^2 + 2 \Phi_{,12} dE^1 dE^2 + \Phi_{,22} (dE^2)^2\right]\, \\
g^{II} & = & \Sigma \left[-\Phi_{,11} (d E^1)^2  + \Phi_{,22} (dE^2)^2\right]\,,
\label{gII2D} \\
g^{III} & = & \beta_1 E^1 \Phi_{,1} \Phi_{,11} (dE^1)^2 + \Sigma \Phi_{,12} dE^1 dE^2
+ \beta_ 2 E^2 \Phi_{,2} \Phi_{,22} (dE^2)^2 \, ,
\label{gIII2D}
\eea
where $\Sigma  =  \beta_1 E^1\Phi_{,1}  + \beta_2 E^2\Phi_{,2} 
$ and 
$\phi_{,a} = \frac{\partial \phi}{\partial E^a}$, etc. 
If $\Sigma = \beta_\Phi  \Phi$, an analysis of the corresponding curvature scalars shows that the singularities are determined  by the zeros of the 
second-order derivatives of $\Phi$, as expressed by the conditions
\be 
\Phi_{,11}\Phi_{,22} -(\Phi_{,12})^2
=0 \ , \ 
\Phi_{,11} \Phi_{,22} 
=0\ , \ \Phi_{,12}= 0\ ,
\label{singrev} 
\ee
which are directly related to the stability conditions and phase transition structure of a system with two thermodynamic degrees of freedom \cite{callen}. 

Notice that the Euler relation is identically satisfied only when all the variables $E^a$ entering the function $\Phi(E^a)$ are allowed to vary. In concrete physical examples, however, it is usually assumed that some of the variables $E^a$ are constant. In such cases, the truncated Euler relation is not necessarily satisfied so that the conformal factor in front of the metrics $g^I_{ab}$ and $g^{II}_{ab}$ as well as in the off-diagonal term of the metric $g^{III}_ {ab} $ cannot be replaced by $\beta_\Phi \Phi $. Then, to obtain explicit expressions for the curvature scalar, it is necessary to choose the coefficients $\beta_a$ in agreement with the quasi-homogeneity conditions.
 We will follow this procedure in the examples in the next section.

\section{Geometrothermodynamics of magnetic materials }
\label{sec:gtdmag}

In this section, we will investigate the main geometric properties 
of the three magnetic models described in Sec. \ref{sec:models}. In each case, we compute explicitly the GTD metrics (\ref{gI2D})-(\ref{gIII2D}), the Riemann curvature tensor, and the curvature scalar, which determine the geometric properties of the corresponding equilibrium space.

\subsection{GTD of the magnetic model without translation }

According to the GTD formalism, the properties of the equilibrium space do not depend on the choice of thermodynamic potential. For comparison with the results presented in Sec. \ref{sec:models}, we choose in this case as thermodynamic potential the magnetization $M$ as a function of the magnetic dipole moment $\mu$, the number of dipoles $N$, the magnetic field $H$, and the temperature $T$. Then, according to Eq.(\ref{mag1}), the fundamental equation is given by 
\be 
M= N  \mu \left[\coth\left(\frac{\mu H}{k_B T} \right)-\frac{k_B T }{ \mu H} \right] .
\ee
Consider now the quasi-homogeneity coefficients of this function, which are important for the explicit computation of the GTD metrics. We see that $M$ is a quasi-homogeneous function, i.e.,  $M(\lambda^{\beta_\mu} \mu , \lambda^{\beta_N} N, \lambda^{\beta_H} H, \lambda^{\beta_T} T ) =  \lambda^{\beta_M} M(\mu,N,H,T)$, if 
the conditions $\beta_T = \beta_\mu + \beta_H $ and $\beta_M = \beta_\mu + \beta_N$ are satisfied. Then, the Euler identity reads $\beta_\mu \mu M_{,\mu} + \beta_N N  M_{,N} + \beta_H H M_{,H} +(\beta_\mu + \beta_H) T M_{,T} = (\beta_\mu + \beta_N) M $. Accordingly, the magnetization depends on four variables, indicating that the equilibrium space is 4-dimensional. However, from a physical point of view it is convenient to keep $N$ and $\mu$ constant so that the magnetization is a function of $H$ and $T$ only, and the equilibrium space is 2-dimensional. Then, using the convention $M = \Phi$, $H=E^1$, and $T=E^2$, the GTD line element (\ref{gI2D}) becomes
\be 
g^I = \Sigma \left( M_{,HH} dH^ 2 + 2 M_{,HT} dH d T+ M_{,TT}dT^2\right)\ ,
\label{gImag1}
\ee
\be
\Sigma = 
\beta_H H M_{,H}
+\beta_T T M_{,T} ,
\ee
from which we can compute the corresponding Riemann curvature tensor $R_{abcd}$ and curvature scalar $R= g^{ac} g^{bd} R_{abcd}$, using the standard approach of differential geometry. In this case, we obtain $R^I=0$, indicating that the equilibrium space is flat. We interpret this result as stating that in this case the metric $g^I_{ab}$  does not contain information about the magnetic model without translations.

We now consider the line element $g^{II}$.  Then, from Eq.(\ref{gdownII}), we obtain
\be
g^{II} = \Sigma 
\left( - M_{,HH} dH^ 2 
+ M_{,TT}dT^2\right)\ .
\label{gIImag1}
\ee
This metric leads to a non-zero curvature scalar, indicating the presence of thermodynamic interaction. In addition, it is easy to show that the curvature  singularities are determined by the roots of the equation
\be
M_{,HH} M_{,TT} = 0 \ ,
\ee
with
\be
M_{,HH} = \frac{2N\mu^3}{k_B^2 T^2}\left[\coth^3\left(\frac{\mu H}{k_B T}\right)
-\coth\left(\frac{\mu H}{k_B T}\right)
-\frac{k_B^3 T^3}{\mu^3 H^3}\right] \ ,
\ee
\be
M_{,TT} = \frac{2\mu^2 H^2 M }{k_B^2 T^ 4}\left[
\coth^2\left(\frac{\mu H}{k_B T}\right) -1
\right]\ .
\ee

\begin{figure}
    \centering
    \includegraphics[scale=0.3]{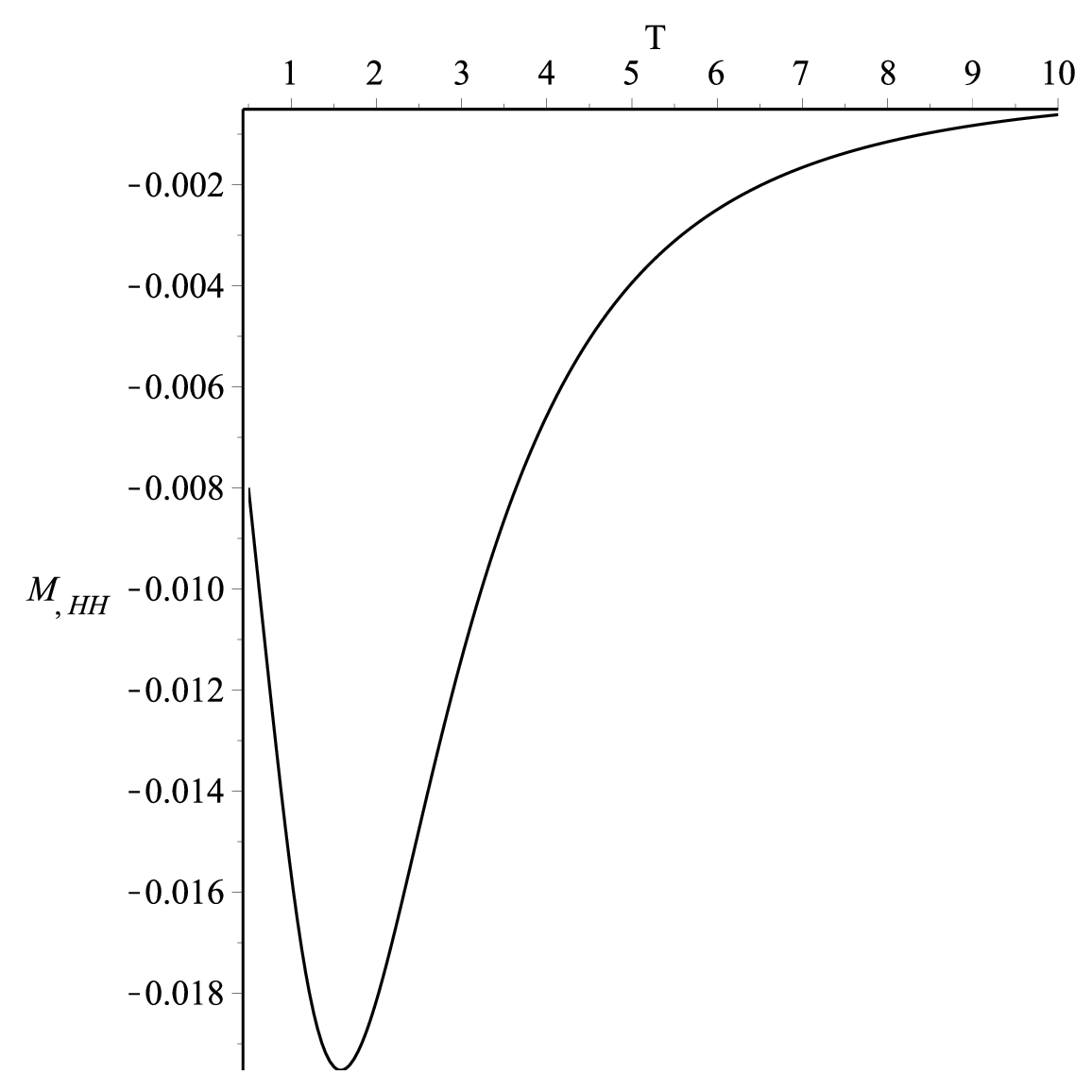}
    \includegraphics[scale=0.3]{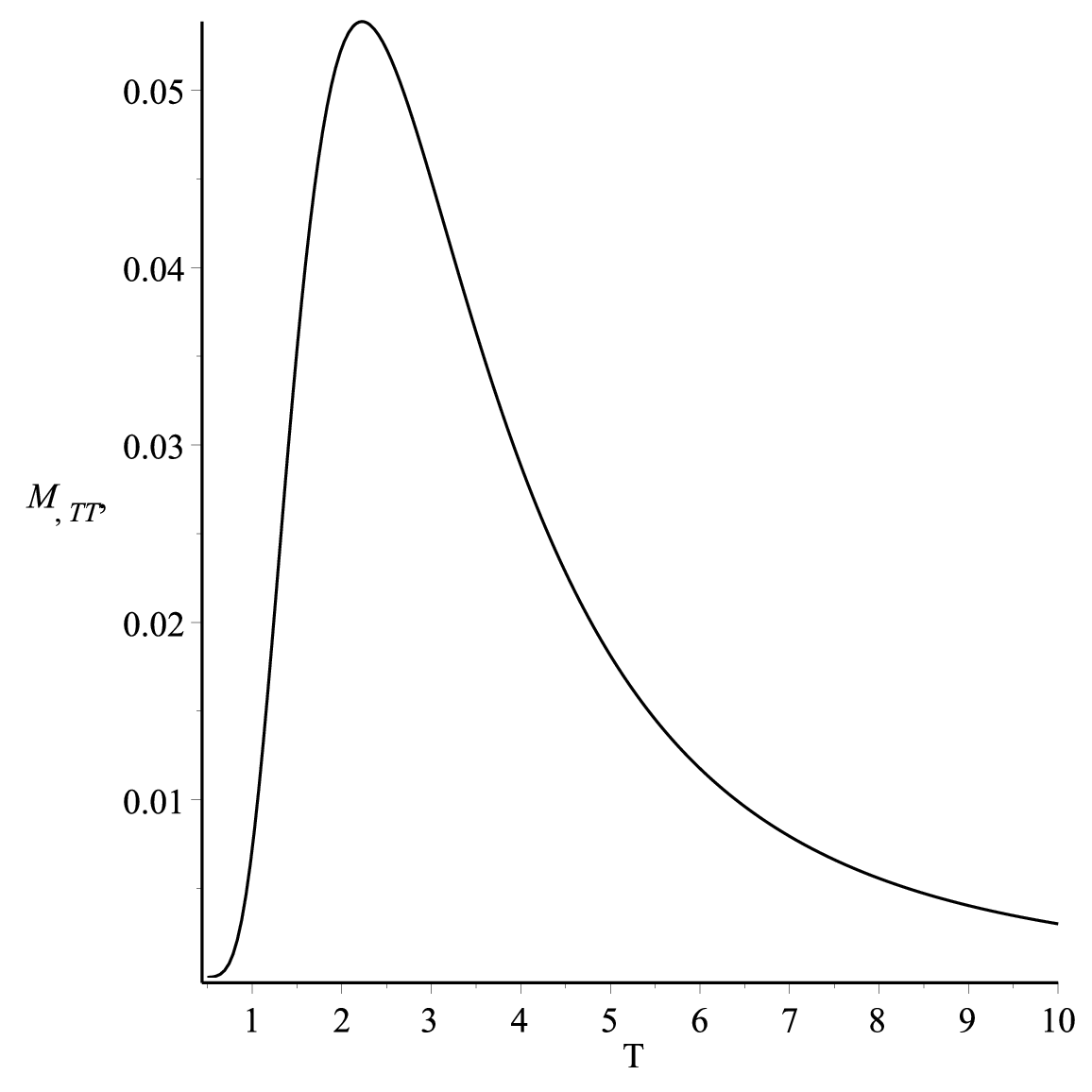}
    \caption{Behavior of the derivatives  $M_{,HH}$ and $M_{,TT}$ as functions of the temperature. For concreteness, we choose $N=1$, $\mu=1$, $k_B=1$ and $H=5$. }
    \label{fign1}
\end{figure}
In Fig.\ref{fign1}, we show the general behavior of  $M_{,HH}$ and $M_{,TT}$ in terms of the temperature. 
We can see that $M_{,HH}$ ($M_{,TT}$) is a smooth negative (positive) function with zeros only in the limits $T\rightarrow 0$ and $T\rightarrow \infty$, which are both excluded from a physical point of view. 
We conclude that the GTD metric $g^{II}_{ab}$ leads to an equilibrium space with a non-zero curvature, which is free of singularities. Thus, this metric captures the thermodynamic interaction due to the presence of the magnetic field in the translation-free dipole system, which turns out to be free of phase transitions. 

\begin{figure}
    \centering
    \includegraphics[scale=0.3]{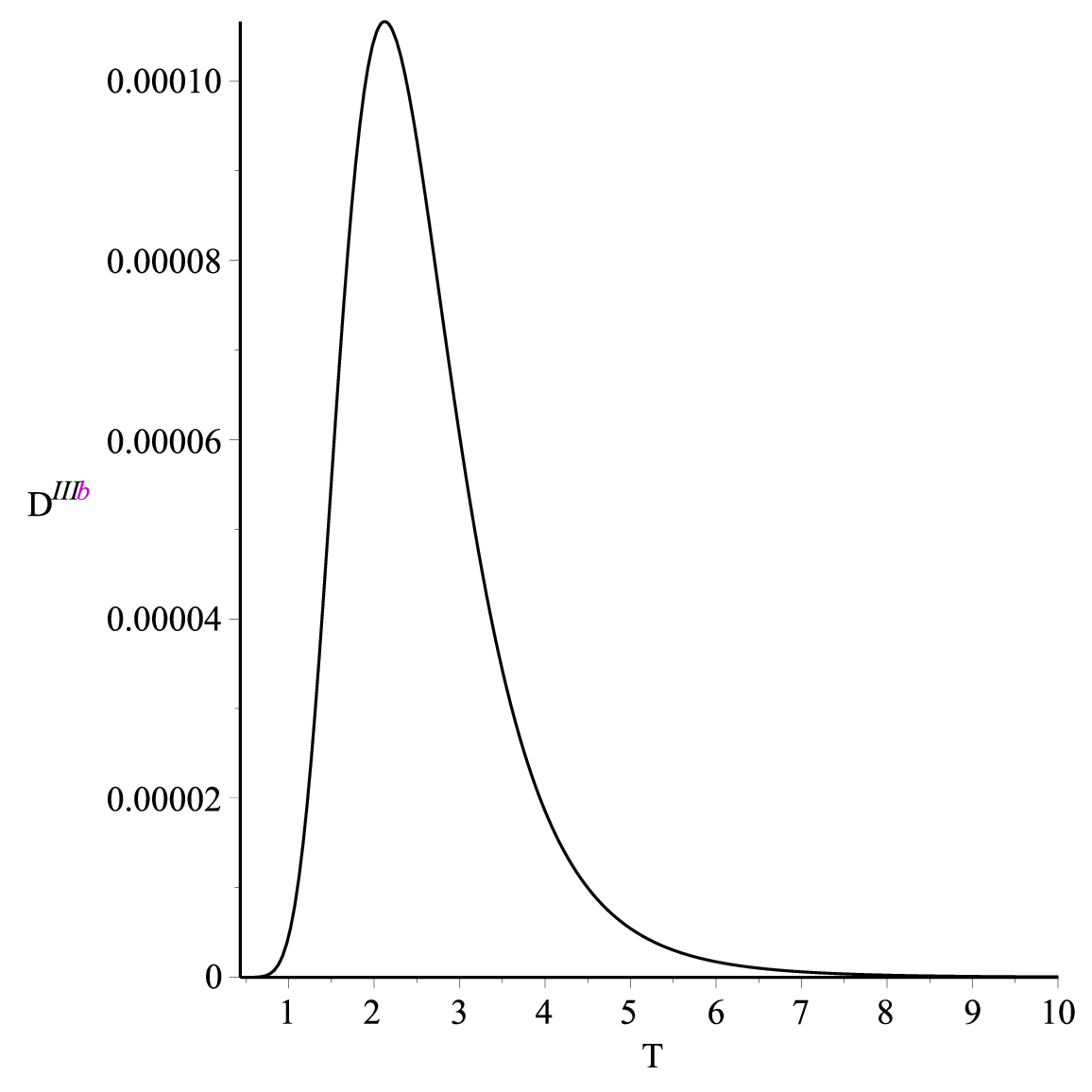}
    \caption{The denominator $D^{III}$ in terms of the temperature. We choose the constants as $k_B=1,\mu=1,N=1,$ and $H=5$. }
    \label{fign2}
\end{figure}

We now consider the metric $g^{III}_{ab}$, which according to Eq.(\ref{gIII2D}) can be expressed as
\be
g^{III}  =  \beta_H H M_{,H} M_{,HH} dH^2 + \Sigma M_{,HT} dH dT 
+ \beta_ T T  M_{,T}M_{,TT} dT ^2  \, ,
\label{gIIImag1}
\ee
One can show that the corresponding scalar curvature can be represented as 
\be 
R^{III} = \frac{N^{III}}{(D^{III})^3}
\label{RIIImag1}
\ee
where $N^{III}$ is a smooth function of its arguments and 
\be 
D^{III}=\beta_H \beta_T H T M_{,H}
M_{,T} M_{,HH} M_{,TT} -\frac{1}{4} \Sigma^2 M_{,HT}^2
\label{DIIImag1}
\ee
so that the zeros of this function determine the curvature singularities. In Fig.\ref{fign2}, we analyze the behavior of the function $D^{III}$. To obtain the explicit form of the plot,  it is necessary to fix the values of the quasi-homogeneity coefficients, taking into account the conditions established above for this fundamental equation. Without loss of generality, we can choose 
$\beta_T=\beta_H$ so that in the resulting final expression the coefficient $\beta_H$ becomes a multiplicative factor, which does not affect the behavior of $D^{III}$. 
The curve of Fig. \ref{fign2} shows that the only zeros are at the asymptotic limits $T\rightarrow 0$ and $T\rightarrow \infty$, implying that in the physically allowed region no zeros exist. We conclude that the curvature scalar, being non-vanishing, shows that thermodynamic interaction is present, but it does not lead to the appearance of phase transitions. 

In this section, we have analyzed the simplest magnetic model without translations from the point of view of the three GTD metrics. As a general result, we find compatible results in the sense that the metrics are able to detect the thermodynamic interaction inside the system and do not predict the presence of phase transitions. This agrees with the results obtained in the previous section, using the approach of statistical physics and classical thermodynamics.

\subsection{GTD of the magnetic model with spin }

In this case, we will consider as fundamental equation the  magnetization $M$ in terms of the temperature (cf. Eq.(\ref{para30})) 
\be 
M = N \mu_B \tanh\left(\frac{\mu_B H}{k_B T}\right) \ ,
\label{feqmag2}
\ee 
where $\mu_B$ represents Bohr's magneton. 

The quasi-homogeneity analysis of the above function shows that the Euler identity can be satisfied only if $\mu_B$  and $k_B$ are considered as thermodynamic variables, i.e., 
$M(\lambda^{\beta_H}H,
\lambda^{\beta_T}T,
\lambda^{\beta_N}N,
\lambda^{\beta_{\mu_B} }\mu_B,
\lambda^{\beta_{k_B} } k_B) = \lambda^
{\beta_M} M$,
if the conditions 
$\beta_M =\beta_N + \beta_{\mu_B} $ and
$\beta_H=\beta_T-\beta_{\mu_B} +\beta_{k_B}  $ are satisfied. The fact that the constants  $\mu_B$  and $k_B$  should be considered as thermodynamic variables opens the possibility of analyzing this model in the context of extended thermodynamics, which in the case of black hole thermodynamic systems has led to the discovery of a new phase transitions \cite{BHchemistry}. In this work, however, we limit ourselves to the physical case in which $N$, $\mu_B$, and $k_B$ are kept constant, implying that we should consider the truncated Euler relation in the form
\be
\beta_H H M{,H} + \beta_T T M_{,T}
= \frac{N\mu_B^2 H (\beta_H-\beta_T)}
{k_B T \cosh^2\left(
\frac{\mu_B H}{k_B T} \right)}
\ .
\ee

In this case, the line elements 
 $g^I$, $g^{II}$, and $g^{III}$ can be represented again as 
(\ref{gImag1}), (\ref{gIImag1}), and (\ref{gIIImag1}), respectively. A direct computation with the fundamental equation (\ref{feqmag2}) shows that the line element $g^I$ corresponds to a flat manifold, indicating that no thermodynamic interaction is present. 

\begin{figure}
    \centering
    \includegraphics[scale=0.3]{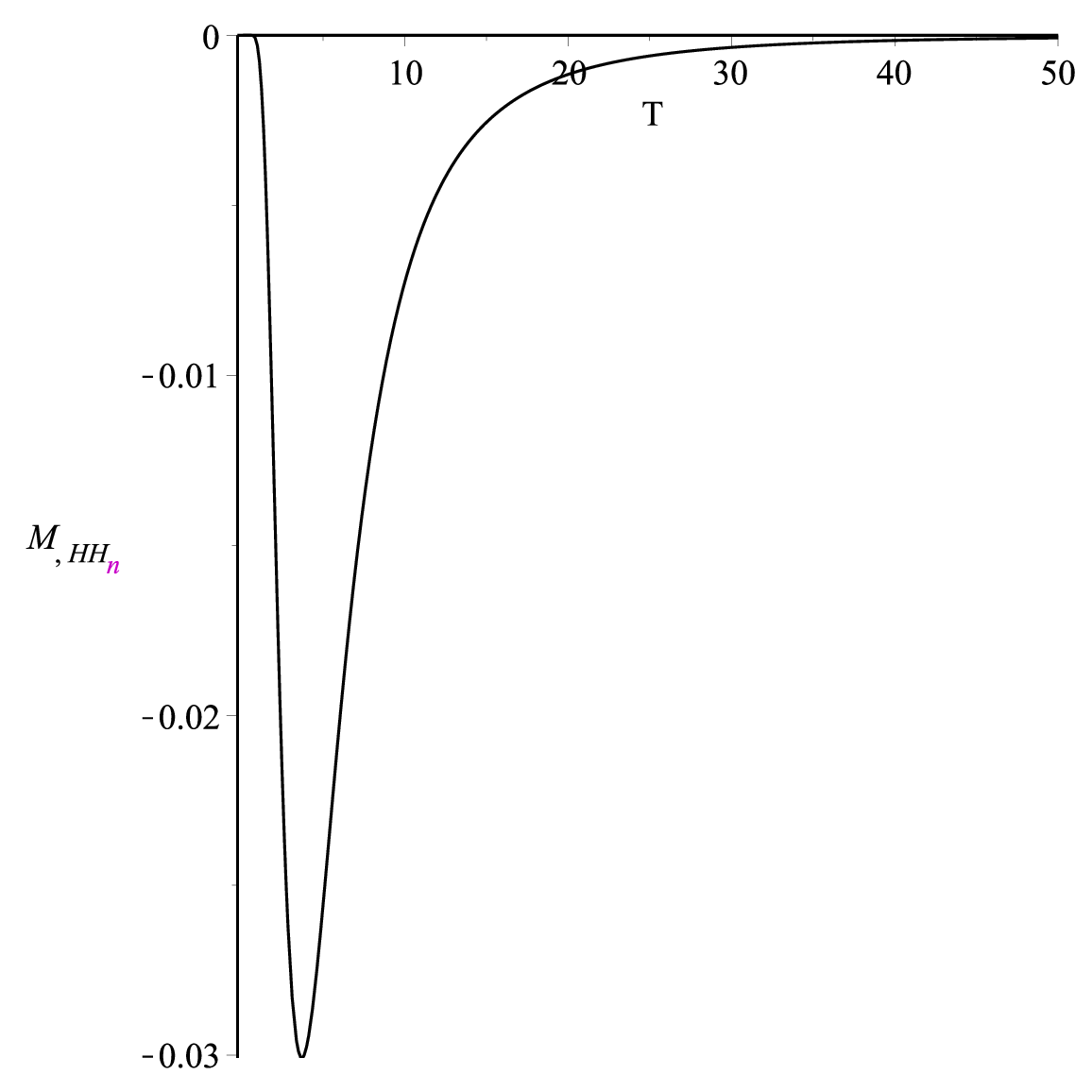}
    \includegraphics[scale=0.3]{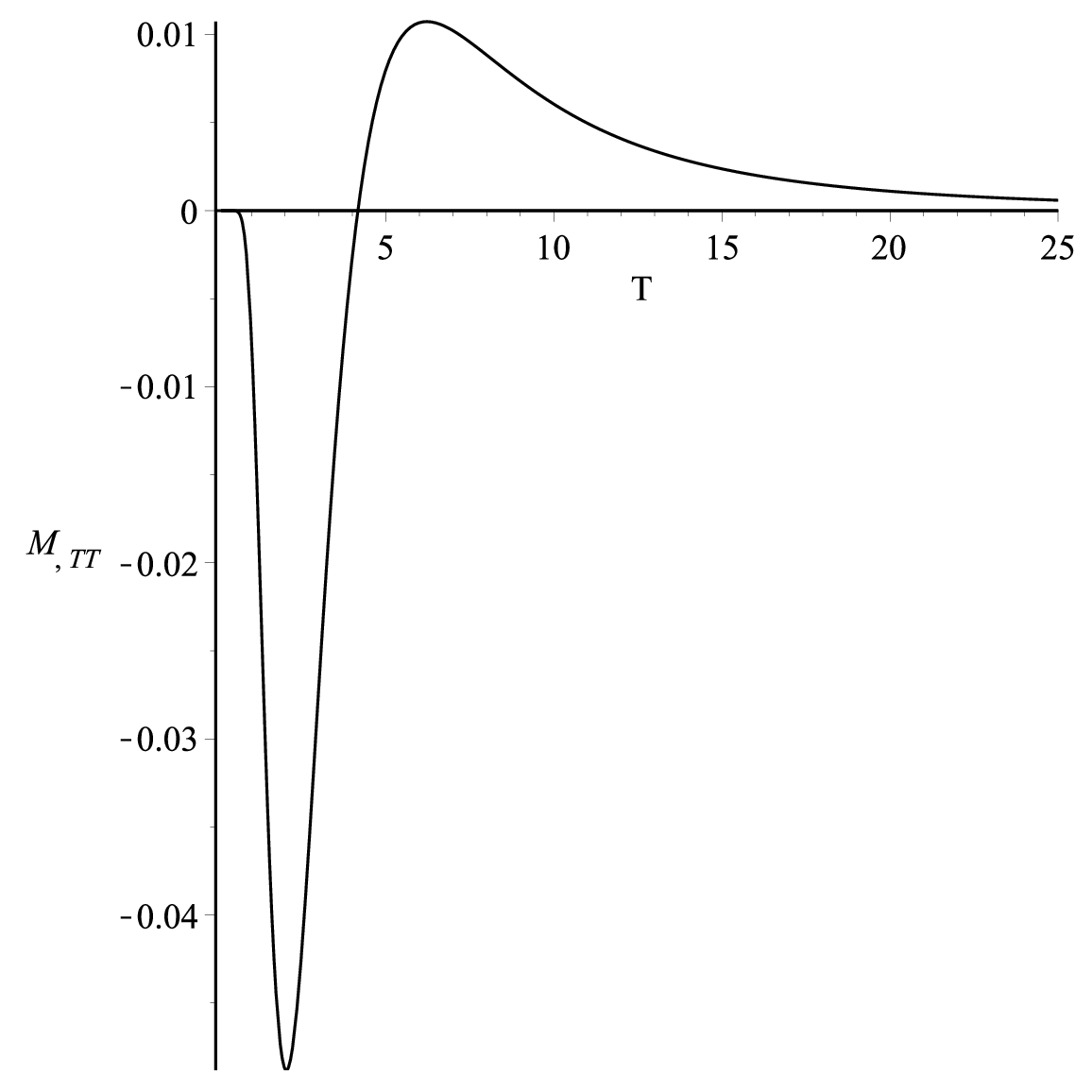}
        \caption{Behavior of the derivatives $M_{,HH}$ and $M_{,TT}$ in terms of the temperature for fixed values of $N$ and $H$.}
    \label{fign3}
\end{figure}

As for the line element $g^{II}$, an analysis of the curvature scalar shows that it becomes singular if the condition $M_{,HH} M_{,TT} = 0$ with
\be
M_{,HH} = \frac{2N\mu_B^3}{k_B^2 T^2}\tanh\left(\frac{\mu_B H}{k_B T}\right)\left[ \tanh^2
\left(\frac{\mu_B H}{k_B T}\right) - 1\right],
\ee 
\be
M_{,TT} = - \frac{2N\mu_B^3 H^2 
\left[\tanh\left(\frac{\mu_B H}{k_B T}\right) - \frac{k_B T}{\mu_B H}\right]
}{k_B^2 T^4\cosh^2\left(\frac{\mu_B H}{k_B T}\right) } .
\label{MTT2}
\ee
In Fig. \ref{fign3}, we see that the function $M_{,HH}$ has no zeros inside the physically allowed range $T\in (0,\infty)$; in contrast, the derivative $M_{,TT}$ possesses a zero inside the allowed region. In fact, this root can be calculated analytically from Eq.(\ref{MTT2}) and the fundamental equation (\ref{feqmag2}), ad we obtain 
\be
MH={Nk_BT}.
\label{singmag2}
\ee
This root indicates the location where the curvature scalar becomes singular, which, according to GTD, corresponds to a second-order phase transition. In Sec.\ref{sec:models}, no phase transition for this case was found by using the heat capacity. However, in general, according to Ehrenfest's scheme, second-order phase transitions correspond to divergences of the response functions \cite{callen}, which can be computed as follows.
The model for a magnetic material with spin is based upon the fundamental equation $M=M(H,T)$, which satisfies the first law of thermodynamics 
\be
dM = I_H dH + I_T dT , \ \ I_H = \frac{\partial M}{\partial H}, \ 
I_T = \frac{\partial M}{\partial T} \ ,
\ee 
where $I_H$ and $I_T$ are the variables dual to $H$ and $T$, respectively
On the other hand, the response functions $C^{ab}$  in classical thermodynamics basically describe how the independent variables in the fundamental equation respond to changes of the dual variables, i.e., 
$C^{ab} = \frac{\partial E^a}{\partial I_b}$. In the magnetic system under study we have that $E^1=H$ and $E^2=T$. Then, the response function 
$C^{22} = \frac{1}{M_{,TT}}$. Therefore, according to Ehrenfest's classification the zeros of $M_{,TT}$ correspond to second-order phase transitions.

We now consider the line element $g^{III}$ given in Eq.(\ref{gIIImag1}). 
The scalar curvature can be represented as in Eqs.(\ref{RIIImag1}) and (\ref{DIIImag1}) so that curvature singularities exist at the locations where the condition $D^{III}=0$  is satisfied. It turns out that by choosing the coefficients of quasi-homogeneity as $\beta_T=\beta_H$, the second term in the expression of $D^{III}$ disappears so that the singularities are determined by the condition $M_{,HH} M_{,TT}=0$, resulting in the singularity (\ref{singmag2}) discussed above. We conclude that the results obtained from the line element $g^{III}$ are compatible with the ones obtained with $g^{II}$.


\subsection{GTD of the mean-field magnetic model }

In this case, to compare our results with those presented in Sec.\ref{sec:models} for the mean-field model, it is convenient to use as thermodynamic potential the generalized  Massieu potential $\tilde S = -F/T$, which is obtained by applying a partial Legendre transformation to the entropy, i.e., $\tilde S = S - U/T$ \cite{callen}. Then, from Eq.(\ref{valormediomagnetico11}),  we get
\be 
\tilde S = \frac{k_{{}_B}}{2}\ln{\left( \frac{4}{1-M^2}\right) }-\frac{q I
M^2}{2T} ,
\label{potencialcampomedio} 
\ee 
so that the independent variables are the magnetization $M$ and the temperature $T$. 
It is interesting to note that this fundamental equation  is a quasi-homogeneous function, i.e.,
$\tilde S (\lambda^{\beta_a}E^a) = 
\lambda ^{\beta_{\tilde S}} \tilde S(E^a)$, only if $k_B$ is a thermodynamic variable with coefficient $\beta_{k_B} = \beta_{\tilde S}=-\beta_T$, and $M$ behaves as an intensive variable, i.e.,  $\beta_M = 0$, which allows us to drastically reduce the explicit form of the GTD metrics, as we will see below. 
As in the case of the magnetic model with spin, the fact that $k_B$ can be considered as a thermodynamic variable would allow us to explore a new phase transition structure as it is done in extended thermodynamics \cite{BHchemistry}.  Here, we will limit ourselves to the investigation of the dependence of the mean-field model from the magnetization and temperature only.

Using the notation $\tilde S = \Phi$, $E^1 = M$, and $E^2 = T$, from Eqs.(\ref{gI2D})-(\ref{gIII2D}), we obtain the following GTD line elements:
\be
g^I =
\frac{\beta_T q I M^2 }{2 T}
\left[
\frac{k_B(1+M^2)}{(1-M^2)^2}dM^2 
-\frac{qI}{T}\left(dM - \frac{M}{T}dT\right)^2 
\right]\ ,
\ee
\be
g^{II} = 
-\frac{\beta_T q I M^2}{2T} \left[
\left(\frac{k_B(1+M^2)}{(1-M^2)^2} - \frac{qI}{T} \right) dM^ 2 +
\frac{qIM^2}{T^3} dT^2
\right ] \ ,
\ee
\be 
g^{III} = \frac{\beta_T q^2 I^2 M^3 }{2T^3}\left( dM dT - \frac{dT^2}{T}\right) \ .
\ee

The calculation of the corresponding curvature scalars is straightforward, and we obtain that the results for each of the above line elements can be represented in a compact form as follows
\be
R^I = - \frac{T}{\beta_T k_B q^2 I^2 M^ 4}\left[ 
2qI (M^4+6M^2+1)(M^2-1) + k_B T (M^2+1)^2
\right]\ ,
\ee 
\be
R^{II} = \frac{k_B T^ 3}{\beta_T q^2 I^2 M^ 4}
\frac{8qIM^2(M^2+3)(M^2-1)-k_BT(M^2+1)^2}
{[ qI (M^2-1)^2 - k_BT (M^2+1)]^2}\ ,
\ee
and
\be
R^{III} = 0\ ,
\ee
respectively.
We can see that $R^I$ and $R^{II}$, being different from zero, indicate the presence of thermodynamic interaction between the dipoles of the system, which can be interpreted as due to the interaction of each spin with its neighbor.  Instead, the metric $g^{III}_{ab}$ turns out to be flat, indicating that a particular aspect of the interaction is absent in the mean-field model. 

Furthermore, the only non-trivial curvature singularity follows from the scalar $R^{II}$ and is given by the roots of the equation $qI (M^2-1)^2 - k_BT (M^2+1)]^2=0$, which can be written as  (cf. Eq.(\ref{temperaturacurie2})) 
\be
T = T_c \frac{(M^2-1)^2}{1+M^2}\ , \ \ T_c = \frac{qI}{k_B}\ ,
\ee
where $T_c$ is the Curie temperature. This result implies that  it is possible to find curvature
singularities at temperature values less than the Curie temperature $T_{{}_c}$, for all
values of magnetization $M$ within the interval
\bea \label{escalaresing2} 0<\frac{(M^2-1)^2}{1+M^2}<1\, .\eea

\begin{figure}[h]
\centering
\includegraphics[scale=0.3]{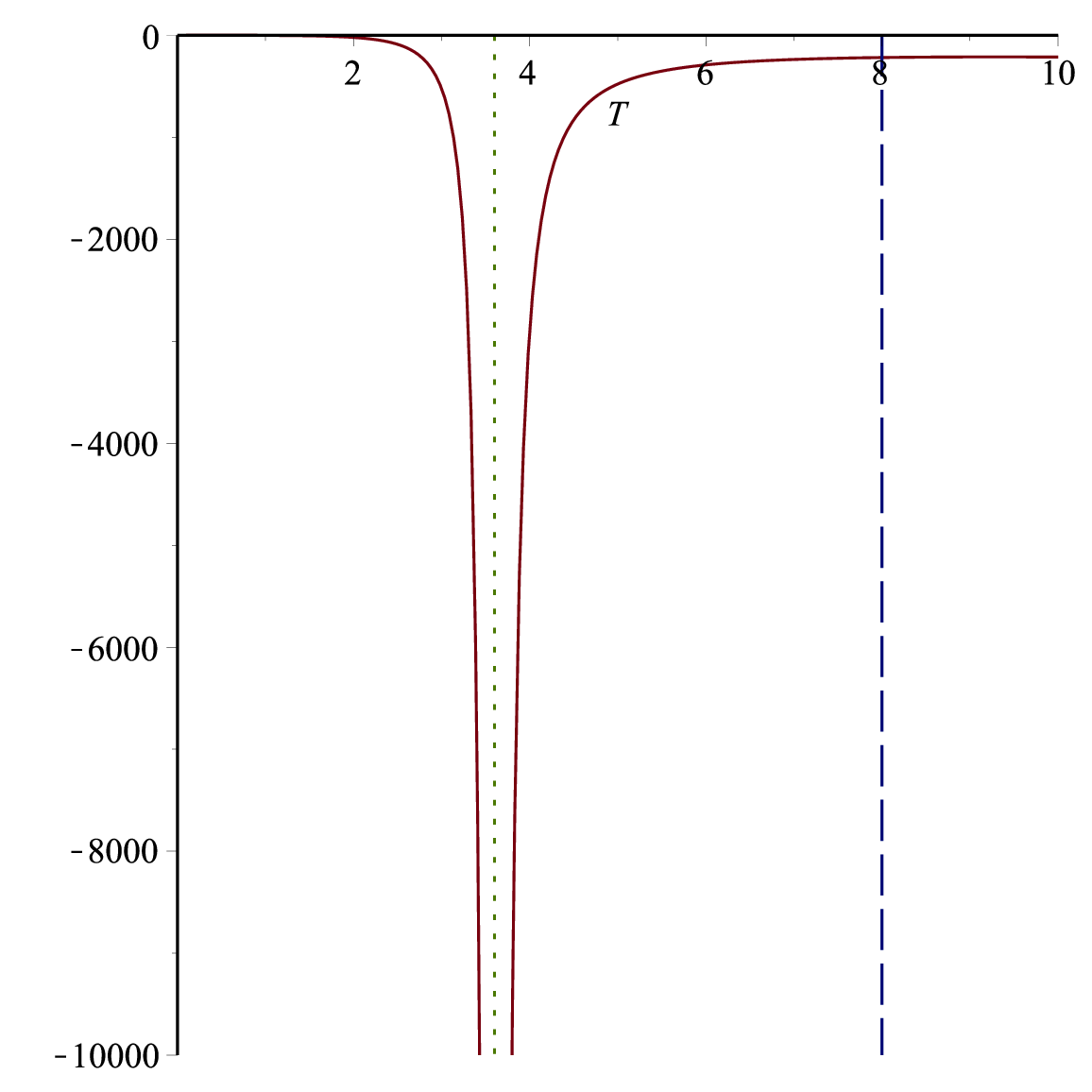} 
\includegraphics[scale=0.3]{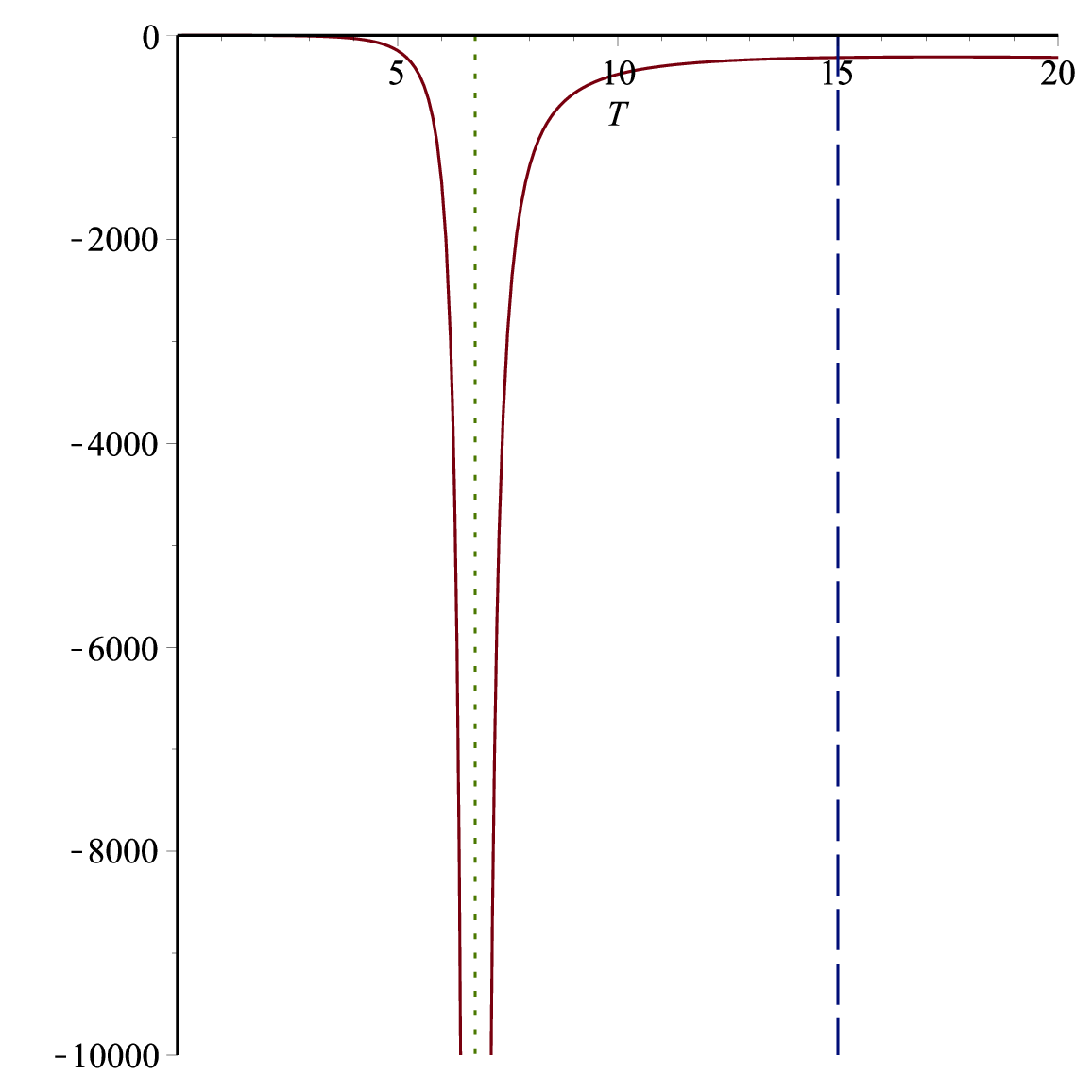}
\caption{
Curvature scalar $R^{II}$ for the mean-field model  as a function of the temperature $T$ in
the interval where singularities exist. 
The solid curve represents the curvature scalar, the dotted line is the singularity, and the dashed line denotes the Curie temperature.
We choose the values $ M=0.5 $ and
$k_{{}_B}=1$  with $q=8$ and $I=1$ (left panel) and with  
$q=3$ and $I=5$ (right panel).
}
\label{fig10}
\end{figure}

Then, we can conclude that the GTD formalism describes
as curvature singularities the phase transitions for temperatures below $T_{{}_c}$, in the same way as indicated by the mean-field magnetic model in statistical physics.
In Fig.\ref{fig10}, we show two specific cases of curvature singularities.

\section{Conclusions}
\label{sec:con}

In this work, we have studied three different models to describe materials with magnetic properties, namely, a simple model based on magnetic dipoles without translations,  a model including the spin of dipoles, and a model based on the mean-field approximation considering interactions between neighboring dipoles. The study has been performed from two points of view, statistical physics and GTD. 

The approach of statistical physics consists in using the properties of each model to construct the corresponding partition functions, from which thermodynamic potentials can be derived by using standard methods \cite{greiner}.  A particular thermodynamic potential that relates the extensive variables of the system is usually considered as the fundamental equation, from which all properties of the system can be derived \cite{callen}. In turn, Legendre transformations can be used to derive alternative fundamental equations, in which other non-extensive variables can appear. An important property of classical thermodynamics is that it is invariant with respect to Legendre transformations, i.e., with respect to the choice of thermodynamic potential.

In this work, we propose to use the formalism of GTD to study magnetic materials. The geometric approaches to thermodynamics consist in endowing the equilibrium space of the system with Riemannian metrics in such a way that thermodynamic properties can be described in terms of geometric concepts. For instance, thermodynamic interaction should be measured by the curvature, and phase transitions should correspond to curvature singularities of the equilibrium space. 
The particular approach of GTD takes into account the invariance of classical thermodynamics with respect to Legendre transformations. To this end, it is necessary to consider only Riemannian metrics, which preserve Legendre invariance. It turns out that there are three different Legendre invariant metrics that can be applied to describe any thermodynamic system. We have found that by imposing the condition of quasi-homogeneity for all thermodynamic systems, consistency is reached for all GTD  metrics in the sense that the three metrics can be applied to describe any particular system simultaneously, leading to compatible results.

In this work, we have applied the quasi-homogeneity condition with the GTD metrics to study some models of magnetic materials. In the case of the simplest model without translations, we found that the metric $g^I_{ab}$ is flat, whereas $g^{II}_{ab}$ and $g^{III}_{ab}$ are curved and free of singularities. We interpret this result as indicating that different metrics take into account different aspects of the thermodynamic interaction. Moreover, the lack of curvature singularities indicates that there are no phase transitions in materials described by this model, a result that corroborates the one obtained by using the statistical approach.

In the case of the model with spin, we obtained that the metric $g^I_{ab}$ is flat, but $g^{II}_{ab}$ and $g^{III}_{ab}$ are non-flat and lead consistently to the appearance of a curvature singularity, which should correspond to a phase transition. Although in the analysis of this model from the point of view of statistical physics, no phase transitions were found by analyzing the behavior of the heat capacity, we proved that, in fact, the phase transition predicted by the GTD approach can be associated with a divergence of a particular response function, which, according to the Ehrenfest classification, corresponds to a second order phase transition. 

In the case of the model based on the mean-field approximation, we proved that the metric $g^{III}_{ab}$ is flat, $g^I_{ab}$ is curved and regular, and $g^{II}_{ab}$ is curved and singular. The analysis of the singularity shows that it corresponds to a second-order phase transition, which occurs when the temperature is less than the Curie temperature. This result coincides with the result obtained using the statistical approach.

In general, we conclude that the GTD approach leads to compatible results in the analysis of all three magnetic models. In the case of the model with spin, GTD predicts the existence of a second-order phase transition that can be physically associated with the divergence of a particular response function. In future works, we expect to further investigate this new phase transition from the point of view of thermodynamics, statistical physics, and possible experimental detection.

Our results show that the GTD metrics can detect thermodynamic interaction and phase transitions in a consistent manner. It seems that different GTD metrics are responsible for different aspects of thermodynamic interaction. To clarify this issue, it will be necessary to perform a detailed analysis of the curvature of each non-flat GTD metric. In this way, we expect to characterize and classify the thermodynamic interaction from a geometric perspective. This problem will be treated in future works.

\section*{Acknowledgements}
The work of MNQ was carried out within the scope of the project CIAS 3750  supported by the Vicerrectoría de Investigaciones de la Universidad Militar Nueva Granada - Vigencia 2022. This work was partially supported by UNAM-DGAPA-PAPIIT, Grant No. 108225.


\begin{thebibliography}{99}


\bibitem{Zemansky} M.W. Zemansky and R.H. Dittman, {\it Heat and Thermodynamics}  (McGraw Hill, New York, 1997).

\bibitem{Hashimoto} K. Hashimoto, 
{\it Possibility of ferromagnetic neutron matter},
Phys. Rev. D {\bf 91}, 085013 (2015). 


\bibitem{Berti} V. Berti , M. Fabrizio , C. Giorgi, {\it A three-dimensional phase
transition model in ferromagnetism: Existence and uniqueness}, J.
Math. Anal. Appl. {\bf 355},  661-674 (2009).


\bibitem{Barman} H. Barman and A. Petrou, {\it Measuring the magnetization of a permanent magnet}, 
Am. J. Phys. {\bf 87}, 275 (2019).


\bibitem{greiner} W. Greiner, L Neise, and H. Stocker, {\em Thermodynamics and Statistical Mechanics} (Springer-Verlag, New York, 1995).


\bibitem{Mohn} P.  Mohn,  {\it Magnetism in the Solid State}
(Springer Verlag, Berlin, 2002). 

\bibitem{Ising} A. Bakk and J. S. Hoye, {\it One-dimensional Ising model applied to
protein folding}, 
 Physica A {\bf 323}, 504 (2003). 
 

\bibitem{Hopfield} J.J. Hopfield, {\em Neural networks and physical systems with emergent collective computational abilities}, Proc. Nat. Acad. Sci. USA {\bf 79} 2554 (1982). 

\bibitem{Amit} D. J. Amit and D. J. Amit, {\em Modeling Brain Function: The World of Attractor Neural Networks}, (Cambridge University Press,  Cambridgeshire, England, 1989).

\bibitem{Sornette} D. Sornette,  {\em Physics and financial economics (1776-2014): puzzles, Ising and agent-based models},
Rep. Prog. Phys. {\bf 77}, 062001 (2014).


\bibitem{Cajueiro} D.O. Cajueiro, {\em Enforcing social behavior in an Ising model with complex
neighborhoods},
Physica A {\bf 390}, 1695 (2011). 

\bibitem{Kohring} G. Kohring, {\it Ising Models of Social Impact: the Role of Cumulative
Advantage},
Journal de Physique I, {\bf 6}, 301 (1996).

\bibitem{Unnar} U. B. Arnalds et al.  {\em A new look on the two-dimensional Ising model: thermal artificial spins},
New J. Phys. {\bf 18}, 023008
(2016).

\bibitem{Kedkanok} K.  Sitarachu and M. Bachmann,  {\em Phase Transitions in the Two-Dimensional Ising
Model from the Microcanonical Perspective}. J. Phys.: Conf. Ser.
 {\bf 1483}
(2020).



\bibitem{Messager} A. Messager and S. Miracle-Sole.{\em Equilibrium States
of the Two-dimensional Ising Model in the Two-Phase Region},
Commun.
Math. Phys. {\bf 40}, 187 (1975).



\bibitem{Vecsei} P. Vecsei,  J. Lado, and C. Flindt,  {\em Lee-Yang theory of the
two-dimensional quantum Ising model},
Phys. Rev. B {\bf 106}, 1 (2022). 

\bibitem{Turban} L. C. Turban,  {\em One-dimensional Ising model with multispin
interactions}. J. Physics A: Mathematical and Theoretical,
{\bf 49},  355002 (2016). 

\bibitem{Dalton}A.  Sakthivadivel, {\em Magnetisation and mean field theory in the Ising
model}, SciPost Physics Lecture Notes {\bf 35} (2022).

\bibitem{castellanos}G. Castellano, {\it  Thermodynamic potentials for simple magnetic systems},  Journal of Magnetism and Magnetic Materials {\bf 260}, 146 (2003).

\bibitem{Barrett} M. Barrett and A. Macdonald, {\it The Form of Magnetic Work in a Fundamental Thermodynamic Equation for a Paramagnet},  Am. J. Phys. {\bf 67},613 (1999).

\bibitem{Berez} V. L. Berezinskii, {\it  Thermodynamics of layered isotropic magnets at low temperatures},
Zh. Eksp. Teor. Fiz. {\bf 64}, 725 (1973).


\bibitem{Apol} M. E. Apol, A. Amadei, and A. Di Nola,  {\it Statistical mechanics and thermodynamics of magnetic and dielectric systems based on magnetization and polarization fluctuations: Application
of the quasiGaussian entropy theory},
J. Chem. Phys. {\bf 116}, 4426 (2002).



\bibitem{Moore} E. E. Moore and A. Perron, {\it  Thermodynamics and Magnetism of SmFe12 Compound Doped with Co and Ni: An Ab Initio Study},  Appl. Sci. {\bf 12}, 4860 (2022).


\bibitem{callen} H.B. Callen, {\em Thermodinamics} (John Wiley \& Sons, Inc., New York, 1981).

\bibitem{Gibbs} J. Gibbs, {\it  Thermodynamics} (Yale University Press, New Haven, CT,
1948).

\bibitem{Charatheodory} C. Charatheodory, {\it Gesammelte Mathematische Werke} (Teubner Verlag, Munich 1995).

\bibitem{Rao} C. R. Rao, {\it  Information and the accuracy attainable in the
estimation of statistical parameters}, 
Bulletin of Calcutta
Mathematical Society {\bf  37}, 81 (1945). 

\bibitem{Amari} S. Amari, {\it Diferential-Geometrical Methods in Statistics}, Lecture Notes in Statistics (Springer Verlag, New York, 2012).


\bibitem{Mrugala1} R. Mrugala, Rep. Math. Phys. {\bf 14}, 419 (1978).

\bibitem{Mrugala2} R. Mrugala, Rep. Math. Phys. {\bf 21}, 197 (1985).

\bibitem{Weinhold} F. Weinhold, J. Chem. Phys. {\bf 63}, 2479 (1975); {\bf 63}, 2484 (1975);
{\bf 63}, 2488 (1975); {\bf 63}, 2496 (1975); {\bf 65}, 558 (1976).

\bibitem{Ruppeiner1} G. Ruppeiner, Phys. Rev. A {\bf 20}, 1608 (1979).

\bibitem{Ruppeiner2} G. Ruppeiner, Rev. Mod. Phys. {\bf 67}, 605 (1995); 68, 313 (1996).

\bibitem{quevedo} H. Quevedo, Geometrothermodynamics, J. Math. Phys. {\bf 48}, 013506 (2007).

\bibitem{quevedo1} H. Quevedo, A. S\'anchez, S. Taj and A. V\'azquez, {\it Phase
Transitions in Geometrothermodynamics}, Gen. Rel. Grav. 
{\bf 43}, 1153 (2011).

\bibitem{quevedo3} H. Quevedo, {\it Geometrothermodynamics of  black holes}, Gen. Rel. Grav. {\bf 40}, 971 (2008). 

\bibitem{quevedo4} J.L. Alv\'arez, H. Quevedo, and A. S\'anchez {\it Unified
geometric description of black hole thermodynamics},
Phys. Rev. D
{\bf 77}, 084004 (2008).

\bibitem{quevedo5} H. Quevedo and A. S\'anchez, {\it Geometrothermodynamics of asymptotically anti-de Sitter black
holes}, JHEP {\bf 09}, 034 (2008).

\bibitem{baxter} R. J. Baxter, {\em Exactly solved models in statistical mechanics} (Academic Press Limited, London, England, 1984).


\bibitem{QSI} H. Quevedo and A. S\'anchez, {\bf Geometric description of BTZ black holes thermodynamics}, Phys. Rev. D {\bf 79}, 024012 (2009).

\bibitem{QSII} H. Quevedo and A. S\'anchez, {\it Geometrothermodinamics of black holes in two dimensions}, Phys. Rev. D {\bf 79}, 087504. (2009).


\bibitem{Arnold} V. I. Arnold, Mathematical Methods of Classical Mechanics
(Springer-Verlag, New York, 1980).


\bibitem{QuevedoMN5} H. Quevedo, M. N. Quevedo, and A. S\'anchez, 
{\em Quasi-homogeneous black hole thermodynamics}, Eur. Phys. J. C {\bf 79}, 229 (2019). 




\bibitem{BHchemistry}
D. Kubiznak and R. M. Mann, {\it Black hole chemistry},  Canadian J. Phys. {\bf 93} 999 (2015).



\end{thebibliography}
\end{document}